\documentclass[rmp,aps,prf,reprint,showpacs,longbibliography]{revtex4-1}

\usepackage[utf8]{inputenc}
\usepackage{graphicx}
\usepackage{amsmath}
\usepackage{amssymb}

\usepackage{graphicx}
\usepackage{epstopdf, epsfig}

\usepackage{color}
\usepackage{amsmath}
\usepackage{amsfonts}
\usepackage{amssymb}
\usepackage{amsbsy}
\usepackage{bm}
\usepackage{graphicx}
\usepackage{extarrows}
\usepackage{tikz}
\usepackage{hyperref}
\usepackage{pgfplots}
\usepackage{ifthen}

\usepackage{tensor}

\newcommand{\ie}{i.\,e.}%
\newcommand{\eg}{e.\,g.}%
\newcommand{\wrt}{w.\,r.\,t.}%
\newcommand{\cf}{cf. }%

\newcommand{\formComma}{\,\text{,}}
\newcommand{\formPeriod}{\,\text{.}}

\newcommand{\R}{\mathbb{R}}

\newcommand{\paraS}{\mathbf{x}} 
\newcommand{\paraTF}{\mathbf{X}}

\newcommand{\TFilm}{\Omega_{h}}

\newcommand{\ucoord}{y^{1}}
\newcommand{\vcoord}{y^{2}}
\newcommand{\ncoord}{\xi}

\newcommand{\veloTFC}{V}
\newcommand{\veloTF}{\boldsymbol{\veloTFC}}
\newcommand{\obveloTFC}{W}
\newcommand{\obveloTF}{\boldsymbol{\obveloTFC}}
\newcommand{\normalVelocity}{v_n}
\newcommand{\relveloTFC}{U}
\newcommand{\relveloTF}{\boldsymbol{\relveloTFC}}

\newcommand{\pTFC}{P}
\newcommand{\pTF}{\boldsymbol{\pTFC}}

\newcommand{\pressTF}{P_{\TFilm}}
\newcommand{\pressure}{p_\surf}

\newcommand{\arbTensorTFC}{T}
\newcommand{\arbTensorTF}{\boldsymbol{\arbTensorTFC}}

\newcommand{\gSC}{g}
\newcommand{\gS}{\boldsymbol{\gSC}}
\newcommand{\gTFC}{G}
\newcommand{\gTF}{\boldsymbol{\gTFC}}

\newcommand{\GammaS}{\Gamma}
\DeclareRobustCommand{\GammaTF}{\text{\raisebox{\depth}{\scalebox{1}[-1]{$\mathbb{L}$}}}}

\newcommand{\tr}{\operatorname{tr}}

\newcommand{\nablaTF}{\nabla}
\newcommand{\dirnablaS}[1]{\boldsymbol{\nabla}^{\!\surf}_{\!#1}}
\newcommand{\dirnablaTF}[1]{\boldsymbol{\nabla}_{\!#1}}

\newcommand{\Lie}[1]{\mathcal{L}_{\!#1}}

\newcommand{\divTF}{\operatorname{div}}

\newcommand{\Tangent}{\mathsf{T}}

\newcommand{\landau}{\operatorname{\mathcal{O}}\!}
\newcommand{\landauNor}{\landau\left( \ncoord \right)}
\newcommand{\landauh}{\landau\left( h^{2} \right)}

\newcommand{\dS}{\textup{d}\surf}

\newcommand{\viscosity}{\nu}
\newcommand{\K}{{\eta}} %
\newcommand{\Kn}{{\omega_n}} %

\newcommand{\atSurf}{\big|_{\surf}}
\newcommand{\AtSurf}{\Big|_{\surf}}

\newcommand{\xb}{\mathbf{x}}
\newcommand{\vbc}{v}%
\newcommand{\vb}{\mathbf{\vbc}}%
\newcommand{\wbc}{w}%
\newcommand{\wb}{\mathbf{\wbc}}%
\newcommand{\pbc}{p}%
\newcommand{\pb}{\mathbf{\pbc}}%
\newcommand{\eb}{\mathbf{e}}%
\newcommand{\mb}{\mathbf{m}}%
\newcommand{\surf}{\mathcal{S}}
\newcommand{\gaussianCurvature}{\kappa}
\newcommand{\Grad}{\nabla}
\newcommand{\Div}{\operatorname{div}}%
\newcommand{\Rot}{\operatorname{rot}}%
\newcommand{\BigRot}{\operatorname{Rot}}%
\newcommand{\DivSurf}{\Div_{\surf}}%
\newcommand{\GradSurf}{\Grad_{\surf}}
\newcommand{\RotSurf}{\Rot_{\surf}}
\newcommand{\BigRotSurf}{\BigRot_{\surf}}
\newcommand{\laplaceBeltrami}{\Delta_{\surf}}
\newcommand{\vecLaplace}{\boldsymbol{\Delta}}
\newcommand{\laplaceDeRham}{\vecLaplace^{\textup{dR}}}

\newcommand{\laplaceDivGrad}{\vecLaplace^{\textup{DG}}}
\newcommand{\LaplaceDeRham}{Laplace-deRham }
\newcommand{\ProjectSurf}{\pi_\surf}
\newcommand{\ProjectPartialTF}{\pi_{\partial\TFilm}^{\flat}}

\newcommand{\surfNormalI}{\nu}
\newcommand{\surfNormal}{\boldsymbol{\surfNormalI}}
\newcommand{\meanCurvature}{\mathcal{H}}
\newcommand{\laplaceDeRhamTilde}{\widehat{\vecLaplace}^{\textup{dR}}}
\newcommand{\laplaceDivGradTilde}{\widehat{\vecLaplace}^{\textup{DG}}}

\newcommand{\extend}[1]{\widehat{#1}}

\newcommand{\vExtC}{\extend{\vbc}}
\newcommand{\vExt}{\extend{\vb}}
\newcommand{\pExtC}{\extend{\pbc}}
\newcommand{\pExt}{\extend{\pb}}
\newcommand{\wExt}{\extend{\wb}}
\newcommand{\triangulation}{{\mathcal{T}}}
\newcommand{\EuBase}[1]{\,\eb^{#1}}
\newcommand{\insertColorbarVertical}[5]{
	\begin{minipage}{1.5cm}
		\begin{flushleft}
			\begin{tikzpicture}
				\node (colorbar) at (0,0) {\includegraphics[width=0.6cm]{colorbar_plain_vertical.png}};
				\draw (0.01,1.4) node {\scriptsize #1};
				\draw (-0.15,0.965) node[anchor=west] {\scriptsize #5};
				\draw (-0.15,0.22666667) node[anchor=west] {\scriptsize #4};
				\draw (-0.15,-0.5116667) node[anchor=west] {\scriptsize #3};
				\draw (-0.15,-1.25) node[anchor=west] {\scriptsize #2};
			\end{tikzpicture}
		\end{flushleft}
	\end{minipage}
}
\newcommand{\insertColorbarHorizontal}[5]{
	\begin{minipage}{4.5cm}
		\begin{center}
			\begin{tikzpicture}
				\node (colorbar) at (0,0) {\includegraphics[width=3.6cm]{colorbar_plain_horizontal.png}};
				\draw (0.0,0.35) node {\scriptsize #1};
				\draw (-1.53,-0.05) node[anchor=north] {\scriptsize #2};
				\draw (-0.53,-0.05) node[anchor=north] {\scriptsize #3};
				\draw (0.47,-0.05) node[anchor=north] {\scriptsize #4};
				\draw (1.47,-0.05) node[anchor=north] {\scriptsize #5};
			\end{tikzpicture}
		\end{center}
	\end{minipage}
}

\newcommand{\surfDiscrete}{{\surf_h}}

\newcommand{\shapeOperatorc}{B}
\newcommand{\shapeOperator}{\mathcal{\shapeOperatorc}}

\newcommand{\weak}[2]{\Big(\hspace{1pt} #1 \hspace{2pt} , \hspace{2pt} #2 \hspace{1pt} \Big)}

\newcommand{\stress}{\boldsymbol{\sigma}}
\newcommand{\ericksenStress}{\stress^{\textup{E}}}
\newcommand{\surfaceEricksenStress}{\ericksenStress_\surf}
\newcommand{\surfaceEricksenStressTilde}{\widehat{\stress}^{\textup{E}}_\surf}
\newcommand{\activeStress}{\stress^{\textup{A}}}
\newcommand{\surfaceActiveStress}{\activeStress_\surf}
\newcommand{\dTau}[1]{\textup{d}_{\tau^m}^{#1}}

\newcommand{\energy}{\mathcal{F}}
\newcommand{\frankOseenEnergy}{\energy^{\mathbf{P}}}
\newcommand{\kineticEnergy}{\energy^{\textup{kin}}}

\newcommand{\tanPen}{\alpha}
\newcommand{\tanPenP}{\tanPen_\pb}
\newcommand{\tanPenV}{\tanPen_\vb}

\newcommand{\hTFC}{H}
\newcommand{\hTF}{\boldsymbol{\hTFC}}
\newcommand{\DTFC}{D}
\newcommand{\DTF}{\boldsymbol{\DTFC}}
\newcommand{\DS}{\DTF_\surf}
\newcommand{\OmegaTFC}{\Omega}
\newcommand{\OmegaTF}{\boldsymbol{\OmegaTFC}}
\newcommand{\OmegaS}{\OmegaTF_\surf}

\newcommand{\hbc}{h}%
\newcommand{\hb}{\mathbf{\hbc}}%

\newcommand{\meshsize}{h_\textup{m}}

\newboolean{submitmode}
\setboolean{submitmode}{true} 

\begin{document}

\title{Hydrodynamic interactions in polar liquid crystals on evolving surfaces} 
\author{Ingo Nitschke}
\affiliation{Institute of Scientific Computing, Technische Universit\"at Dresden, Germany}
\author{Sebastian Reuther}
\affiliation{Institute of Scientific Computing, Technische Universit\"at Dresden, Germany}
\author{Axel Voigt}
\affiliation{Institute of Scientific Computing, Technische Universit\"at Dresden and Dresden Center for Computational Materials Science (DCMS) and Center for Systems Biology Dresden (CSBD), Dresden, Germany}

\begin{abstract}
We consider the derivation and numerical solution of the flow of passive and active polar liquid crystals, whose molecular orientation is subjected to a tangential anchoring on an evolving curved surface. 
The underlying passive model is a simplified surface Ericksen-Leslie model, which is derived as a thin-film limit of the corresponding three-dimensional equations
with appropriate boundary conditions. A finite element discretization is considered and the effect of hydrodynamics on the interplay of topology, geometric 
properties and defect dynamics is studied for this model on various stationary and evolving surfaces. 
Additionally, we consider an active model. We propose a surface formulation for an active polar viscous gel and exemplarily demonstrate the effect of the underlying curvature on the location of topological defects on a torus.
\end{abstract}


\maketitle

\section{Introduction}
\label{sec1}
Liquid crystals (LCs) are partially ordered materials that combine the fluidity of liquids with the orientational order of 
crystalline solids \cite{deGennesetal_book_1993,Chaikinetal_book_1995}. Topological defects are a key feature of LCs if considered under external constraints. In particular on curved 
surfaces these defects are important and have been intensively studied on a sphere \cite{Dzubiellaetal_PRE_2000,Batesetal_SM_2010,Shin_PRL_2008,Dhakaletal_PRE_2012,Koningetal_SM_2013,Napolietal_PRL_2012} 
and under more complicated constraints \cite{Stark_PR_2001,Prinsenetal_PRE_2003,Martinezetal_NM_2014}. LCs on curved surfaces can be realized on various levels. 
One possibility is to prepare a double emulsion of two concentric droplets \cite{Fernandeznievesetal_PRL_2007} for which the intervening shell is 
filled with molecular or colloidal LCs which show planar anchoring at the two curved interfaces \cite{Lopezleonetal_JPCM_2012,Liangetal_PTRSA_2013,Liangetal_PRL_2011}. 
Also air bubbles covered by microrods 
have been prepared and studied in real-space \cite{Zhouetal_ACIE_2008}. Moreover, topological defects for charged colloidal spheres confined on a sphere 
were experimentally studied \cite{Guerraetal_N_2018}. Ellipsoidal colloids bound to curved fluid-fluid interfaces with negative 
Gaussian curvature \cite{Liuetal_ARCMP_2018} and spherical droplets covered with aspherical surfactants \cite{Yangetal_N_2018} were 
explored. Even living and motile ``particles" like cells \cite{Badeetal_BJ_2017} and suspensions of microtubules and kinesin 
\cite{Keberetal_Science_2014,Ellisetal_NP_2018} were recently studied on surfaces with non-constant curvature. In all these studies
a tight coupling between topology, geometric properties and defect dynamics is observed. 
In equilibrium defects are positioned according to geometric properties of the surface \cite{Lubenskyetal_JP_1992,Nelson_NL_2002,Kraljetal_SM_2011}.
Creation and annihilation of defects can result from geometric interaction, leading to different realizations of the Poincar\'e-Hopf theorem on topologically equivalent but 
geometrically different surfaces \cite{Nestleretal_JNS_2018}.
Also changes in the phase diagram can be induced by the geometry, \eg\ allowing for
coexistence of isotropic and nematic phases in surface LCs \cite{Nitschkeetal_PRSA_2018}. 
In active systems the observed phenomena are even richer, including, \eg, oscillating defect patterns 
\cite{Keberetal_Science_2014,Alaimoetal_SR_2017} and circulating band structures \cite{Sknepneketal_PRE_2015}. The effect of hydrodynamics on these phenomena is more or less unexplored.  

Most of the theoretical studies of these phenomena use particle methods. Despite the interest in such methods a continuous
description would be more essential for predicting and understanding the macroscopic relation between type and position 
of the defects and geometric properties of the surface. Also the influence of hydrodynamics and dynamic shape changes on these relations would be
much more appropriate to study within a continuous approach. However, a coherent model, which accounts for the complex
interplay between topology, geometry, defect interactions, hydrodynamics and shape changes, is still lacking. 
In \cite{Napolietal_PRE_2016}, an attempt in this direction is proposed but for a fixed surface.
We here extend this approach and propose a minimal continuous surface hydrodynamic 
LC model, which contains the evolution of the surface, tangential polar ordering and surface hydrodynamics. The passive model is derived
as a thin-film limit of the simplified Ericksen-Leslie model \cite{Linetal_ARMA_2000}. We describe a numerical approach to solve this model
on general surfaces and demonstrate by simulations various expected and some unexpected phenomena on ellipsoidal and toroidal surfaces.
These phenomena result from the tight coupling of the geometry with the fluid velocity and the director field. However, a full exploration of the rich nonlinear phenomena resulting from these relations goes beyond the scope of the paper. This also holds for the extension to active systems.
The proposed model of surface active polar viscous gels follows as a thin-film limit of a three-dimensional active polar viscous gel model, which 
combines a more general Ericksen-Leslie model with active components \cite{Simhaetal_PRL_2002,Kruseetal_PRL_2004}. The model can be derived and numerically solved using the same concepts. We here only formulate the model and exemplarily demonstrate numerically the effect 
of the underlying curvature on the location of topological defects in an active system. Throughout the whole paper we consider the evolution of the surface to be prescribed and the surface to be decoupled from the surrounding bulk phases in order to highlight the surface hydrodynamics and its coupling with topological and geometric effects.  

\section{The Ericksen-Leslie model}
\label{sec2}
The Ericksen-Leslie model \cite{Ericksen_JR_1961,Ericksen_ALC_1975,Leslie_ARMA_1968} is an established model 
for LCs, whose relaxation dynamics are affected by hydrodynamics. In \cite{Linetal_ARMA_2000} 
a simplified model was introduced and analyzed. This system already retains the main properties of the original 
Ericksen-Leslie model \cite{Linetal_ARMA_2010,Huetal_CMS_2013,Wangetal_ARMA_2013,Huangetal_CMP_2014}
and will be considered as a starting model to derive a surface hydrodynamic LC model by means of a thin-film limit, see appendix \ref{app:thin_film_limit}. The resulting simplified surface Ericksen-Leslie model (cf. eqs. \eqref{eqNSSAlt} -- \eqref{eqDirSAlt}) reads
\begin{align}
	\label{eq1}
	\ProjectSurf\partial_{t}\vb + \dirnablaS{\vb}\vb &= \normalVelocity\left( \shapeOperator\vb + \GradSurf\normalVelocity \right) - \GradSurf\pressure -\viscosity\laplaceDeRham\vb \notag \\
	&\hspace{0.5cm} + \viscosity\left( 2\gaussianCurvature\vb + \GradSurf\left(\normalVelocity\meanCurvature\right) - 2\DivSurf\left(\normalVelocity\shapeOperator\right) \notag \right) \\
	&\hspace{0.5cm} - \lambda\DivSurf\surfaceEricksenStress \\
	\label{eq2}
	\DivSurf\vb &= \normalVelocity\meanCurvature\\
	\label{eq3}
	\ProjectSurf\partial_{t}\pb + \dirnablaS{\vb}\pb &= \K\left( \laplaceDivGrad\pb - \shapeOperator^{2}\pb \right) \notag \\
	&\hspace{0.5cm} - \Kn\left( \left\| \pb \right\|_{\surf}^{2} - 1 \right)\pb
\end{align}
where $\vb(t) \in \Tangent\surf(t)$ denotes the tangential surface velocity, $\pb(t) \in \Tangent \surf(t)$ the tangential director field, representing the averaged molecular orientation, $\pressure(\xb,t) \in \R$ the surface pressure and $\surfaceEricksenStress = \left(\GradSurf\pb\right)^{T} \GradSurf\pb + \left( \shapeOperator\pb \right) \otimes \left( \shapeOperator\pb \right)$ the extrinsic surface Ericksen stress tensor. 
The model is defined on a compact smooth Riemannian surface $\surf(t)$. We consider initial conditions 
$ \vb \left( \xb, t=0  \right) = \vb_{0}(\xb) \in \Tangent_{\xb}\surf(0) $ and 
$ \pb \left( \xb, t=0  \right) = \pb_{0}(\xb) \in \Tangent_{\xb}\surf(0) $. The positive constants $\viscosity$, $\lambda$ and $\K$ denote the fluid viscosity, the competition between kinetic and elastic potential energy and the elastic relaxation time for the molecular orientation field, respectively. $\gaussianCurvature$ is the Gaussian curvature,  $\meanCurvature$ the mean curvature, $\shapeOperator$ the shape operator, $\normalVelocity$ a prescribed normal velocity of the surface and $\Kn$ a penalization parameter to enforce $\| \pb \| = 1$ weakly. 
$\Tangent_{\xb}\surf(t)$ is the tangent space on $ \xb \in \surf(t)$, $\Tangent \surf(t) = \sqcup_{\xb \in \surf(t)} \Tangent_{\xb} \surf(t)$ the tangent bundle, 
$\ProjectSurf$ the projection to the tangential space \wrt\ the surface $\surf(t)$ and $\dirnablaS{\vb}, \GradSurf, \DivSurf$, $\laplaceDeRham$ as well as $\laplaceDivGrad$ the covariant directional derivative, covariant gradient, surface divergence, \LaplaceDeRham operator and Bochner Laplacian, respectively. 
The system combines an incompressible surface Navier-Stokes equation \cite{Reutheretal_MMS_2015,Yavarietal_JNS_2016,Jankuhnetal_preprint_2017,Miura_arXiv_2017} with a weak surface Frank-Oseen model \cite{Nestleretal_JNS_2018} on an evolving surface. For a general discussion on transport of vector-valued quantities on evolving surfaces we refer to \cite{NitschkeVoigt_2019}. The used formulation with the projection operator $\ProjectSurf$ requires the presence of an embedding space, which is $\R^3$ in our case, see appendix \ref{app:thin_film_limit} for details. 

For $\lambda = 0$ eqs. \eqref{eq1} and \eqref{eq2} are the surface Navier-Stokes equation for an incompressible surface fluid on an evolving surface. These equations can be obtained as a thin-film limit of the three-dimensional Navier-Stokes equation in an evolving domain \cite{Miura_arXiv_2017} or by a variational derivation \cite{Yavarietal_JNS_2016}. If only a stationary surface is considered, \ie\ $\normalVelocity = 0$, the equations reduce to the incompressible surface Navier-Stokes equation as considered in \cite{Ebinetal_AM_1970,Mitreaetal_MA_2001,Arroyoetal_PRE_2009,Nitschkeetal_JFM_2012,Reutheretal_MMS_2015,Nitschkeetal_book_2017,Reutheretal_PF_2018}. Compared with its counterpart in flat space, not only the operators are replaced by the corresponding surface operators, also an additional contribution from the Gaussian curvature arises. This additional term results from the surface divergence of the surface strain rate tensor, see \cite{Arroyoetal_PRE_2009,Jankuhnetal_preprint_2017}. The unusual sign results from the definition of the surface \LaplaceDeRham operator \cite{Abrahametal_Springer_1988}. Eq. \eqref{eq3} with $\vb = \mathbf{0}$, $\normalVelocity = 0$ and the \LaplaceDeRham operator $\laplaceDeRham$ instead of the Bochner Laplacian $\laplaceDivGrad$ has been derived as a thin-film limit in \cite{Nestleretal_JNS_2018} and models the $L^2$-gradient flow of a weak surface Frank-Oseen energy. The different operators result from different one-constant approximations in the Frank-Oseen energy, see appendix \ref{app:thin_film_limit} for details. Again an additional geometric term enters in this equation if compared with the corresponding model in flat space. The term with the shape operator $\shapeOperator$ results from the influence of the embedding \cite{Napolietal_PRL_2012,Segattietal_PRE_2014,Nestleretal_JNS_2018}. The coupled system eqs. \eqref{eq1} - \eqref{eq3} with $\normalVelocity = 0$ can be considered as the surface counterpart of the model in \cite{Linetal_ARMA_2000}. Related surface models have been proposed and analyzed in \cite{Shkoller_CPDE_2002,Napolietal_PRE_2016}. The model in \cite{Shkoller_CPDE_2002} is derived from a variational principle on a stationary surface and thus only contains intrinsic terms. It differs from eqs. \eqref{eq1} - \eqref{eq3} with $\normalVelocity = 0$ by the extrinsic term $\shapeOperator^2 \pb$ and the extrinsic contribution in the surface Ericksen stress tensor $\left( \shapeOperator\pb \right) \otimes \left( \shapeOperator\pb \right)$. The model in \cite{Napolietal_PRE_2016} coincides with our formulation with $\normalVelocity = 0$ if a specific parameter set is considered, see also \cite{Napolietal_JP_2010}. However, note that in their notation the symbol $\tilde{\Grad}_{\surf}$ denotes the surface gradient operator, while we use $\GradSurf$ as the covariant gradient operator. Both are related to each other by $\GradSurf\pb + \surfNormal \otimes \shapeOperator \pb = \tilde{\Grad}_{\surf} \pb$, where $\surfNormal$ denotes the surface normal. 

\section{Numerical method}
\label{sec3}
Eqs. \eqref{eq1} - \eqref{eq3} are a system of vector-valued surface PDEs. Numerical approaches have been developed for such equations on general surfaces only recently, see \cite{Reutheretal_PF_2018,Olshanskiietal_arXiv_2018} for the surface (Navier-)Stokes equation, \cite{Nestleretal_JNS_2018} for the surface Frank-Oseen model and \cite{Hansboetal_arXiv_2016} for a surface vector-Laplace equation. Earlier approaches using vector spherical harmonics, \eg\ \cite{Freedenetal_book_2009,Nestleretal_JNS_2018,Praetoriusetal_PRE_2018}, are restricted to a sphere or radial manifold shapes \cite{Grossetal_JCP_2018} and approaches which rewrite the surface Navier-Stokes equation in a surface vorticity-stream function formulation \cite{Nitschkeetal_JFM_2012,Reutheretal_MMS_2015,Mickelinetal_PRL_2018,Reusken_preprint_2018} are limited to surfaces with genus $g(S) = 0$, see \cite{Nitschkeetal_book_2017,Reutheretal_PF_2018} for details. For the numerical solution of eqs. \eqref{eq1} - \eqref{eq3} we combine the methods in \cite{Reutheretal_PF_2018,Nestleretal_JNS_2018} in an operator splitting approach. The idea behind these methods is to extend the variational space from vectors in $\Tangent \surf$ to vectors in $\R^3$, while penalizing the normal components. This allows to split the vector-valued surface PDE into a set of coupled scalar-valued surface PDEs for each component for which established numerical methods are available, see the review \cite{Dziuketal_AN_2013}. 

The corresponding extended problem to eqs. \eqref{eq1} - \eqref{eq3} reads
\begin{align}
	\label{eq4}
	\ProjectSurf\partial_{t}\vExt + \dirnablaS{\vExt}\vExt &= \normalVelocity\left( \shapeOperator\vExt + \GradSurf\normalVelocity \right) - \GradSurf\pressure -\viscosity\laplaceDeRhamTilde\vExt \notag \\
	&\hspace{0.5cm} + \viscosity\left( 2\gaussianCurvature\vExt + \GradSurf\left(\normalVelocity\meanCurvature\right) - 2\DivSurf\left(\normalVelocity\shapeOperator\right) \notag \right) \\
	&\hspace{0.5cm} - \lambda\DivSurf\surfaceEricksenStressTilde - \tanPenV (\vExt \cdot \surfNormal) \surfNormal \\
	\label{eq5}
	\DivSurf\vExt &= \normalVelocity\meanCurvature\\
	\label{eq6}
	\ProjectSurf\partial_{t}\pExt + \dirnablaS{\vExt}\pExt &= \K\left( \laplaceDivGradTilde\pExt - \shapeOperator^{2}\pExt \right) - \Kn\left( \left\| \pExt \right\|^{2} - 1 \right)\pExt \notag \\
	&\hspace{0.5cm} - \tanPenP \left(\surfNormal  \cdot \pExt \right) \surfNormal 
\end{align}
with $ \vExt = \vExtC_{x}\EuBase{x} + \vExtC_{y}\EuBase{y} + \vExtC_{z}\EuBase{z}$, $\pExt = \pExtC_{x}\EuBase{x} + \pExtC_{y}\EuBase{y} + \pExtC_{z}\EuBase{z} \in \R^3  $ and $\surfaceEricksenStressTilde = \left(\GradSurf\pExt\right)^{T} \GradSurf\pExt + \left( \shapeOperator\pExt \right) \otimes \left( \shapeOperator\pExt \right)$.
We further use $\DivSurf \vExt = \nabla \cdot \vExt - \surfNormal \cdot (\nabla \vExt \cdot \surfNormal)$, $\RotSurf \vExt = -\DivSurf (\surfNormal \times \vExt)$ and $\laplaceDivGradTilde\pExt = \DivSurf\GradSurf\pExt$ and $\laplaceDeRhamTilde \vExt = - (\RotSurf\RotSurf\vExt - \GradSurf (\normalVelocity \meanCurvature))$ since $\DivSurf \vExt = - \normalVelocity \meanCurvature$. 
The normal components $\vExt \cdot \surfNormal$ and $\pExt \cdot \surfNormal$ are penalized by the additional terms $\tanPenV (\surfNormal \cdot \vExt) \surfNormal$ and $\tanPenP (\surfNormal \cdot \pExt) \surfNormal$ with penalization parameters $\tanPenV$ and $\tanPenP$.
For convergence results in $\tanPenV$ and $\tanPenP$ for the surface Navier-Stokes and the surface Frank-Oseen problem we refer to \cite{Reutheretal_PF_2018,Nestleretal_JNS_2018}. 
Without these penalization terms the system of equations \eqref{eq4} - \eqref{eq6} is an under-determined problem, since the vector fields are considered in $\R^3$ and therefore the normal components are completely arbitrary, see \cite{Reutheretal_PF_2018,Nestleretal_JNS_2018} for details.
Eqs. \eqref{eq4} - \eqref{eq6} can now be solved for each component $\vExtC_x$, $\vExtC_y$, $\vExtC_z$, $\pExtC_x$, $\pExtC_y$, $\pExtC_z$ and $\pressure$ using standard approaches for scalar-valued problems on surfaces, such as the surface finite element method \cite{DziukElliott_JCM_2007,DziukElliott_IMAJNA_2007,Dziuketal_AN_2013}, level set approaches \cite{Bertalmioetal_JCP_2001,Greeretal_JCP_2006,Stoeckeretal_JIS_2008,Dziuketal_IFB_2008} or diffuse interface approximations \cite{Raetzetal_CMS_2006}. We consider a simple operator splitting approach and solve eqs. \eqref{eq4} - \eqref{eq5} and eq. \eqref{eq6} iteratively in each time step, employing the same surface finite element discretizations as in \cite{Reutheretal_PF_2018,Nestleretal_JNS_2018}. A semi-implicit Euler discretization in time is used. Thereby, the nonlinear transport term in eq. \eqref{eq4} and the norm-$1$ penalization term in eq. \eqref{eq6} are linearized in time by a Taylor-$1$ expansion and the transport term in eq. \eqref{eq6} as well as the term including the surface Ericksen stress tensor in eq. \eqref{eq4} are coupling terms in the operator splitting scheme. Additionally, we employ an adaptive time-stepping scheme which is based on the combination of changes in the surface Frank-Oseen energy and the Courant-Friedrichs-Lewy (CFL) condition. For more details we refer to appendix \ref{app:numerics}. The resulting discrete equations are implemented in the FEM-toolbox AMDiS \cite{VeyVoigt_CVS_2007,Witkowskietal_ACM_2015}, where we additionally use a domain decomposition ansatz to efficiently distribute the workload on many cores systems. 

\section{Results}
In the following simulations we use $\lambda=0.5$, $\tanPenV=10^{2}$, $\Kn=10^{2}$ and $\tanPenP=10^{5}$, where all parameters are treated as nondimensional. We compare the solution of eqs. \eqref{eq4} - \eqref{eq6} (the so-called \textit{wet} case) and the solution of eq. \eqref{eq6} with $\vExt=0$ (the so-called \textit{dry} case). To highlight the differences we take the surface Frank-Oseen energy $\frankOseenEnergy$ and the surface kinetic energy $\kineticEnergy$ into account, which read in the extended form incorporating the penalization term, 
\begin{align*}
	\frankOseenEnergy &:= \int_\surf\frac{\K}{2}\left( \|\GradSurf\pExt\|^2 + \left(\shapeOperator\pExt\right)^2 \right) + \frac{\Kn}{4}\left(\|\pExt\|^2-1\right)^2\dS \\
	&\qquad+ \frac{\tanPenP}{2}\int_\surf\left(\pExt\cdot\surfNormal\right)^2\dS \\
	\kineticEnergy &:= \frac{1}{2}\int_\surf\vExt^2\dS \formPeriod
\end{align*}

\begin{figure}[t!]
	\centering
	\ifthenelse{\boolean{submitmode}}{
		\begin{minipage}{0.48\textwidth}
			\centering
			\includegraphics[width=\textwidth]{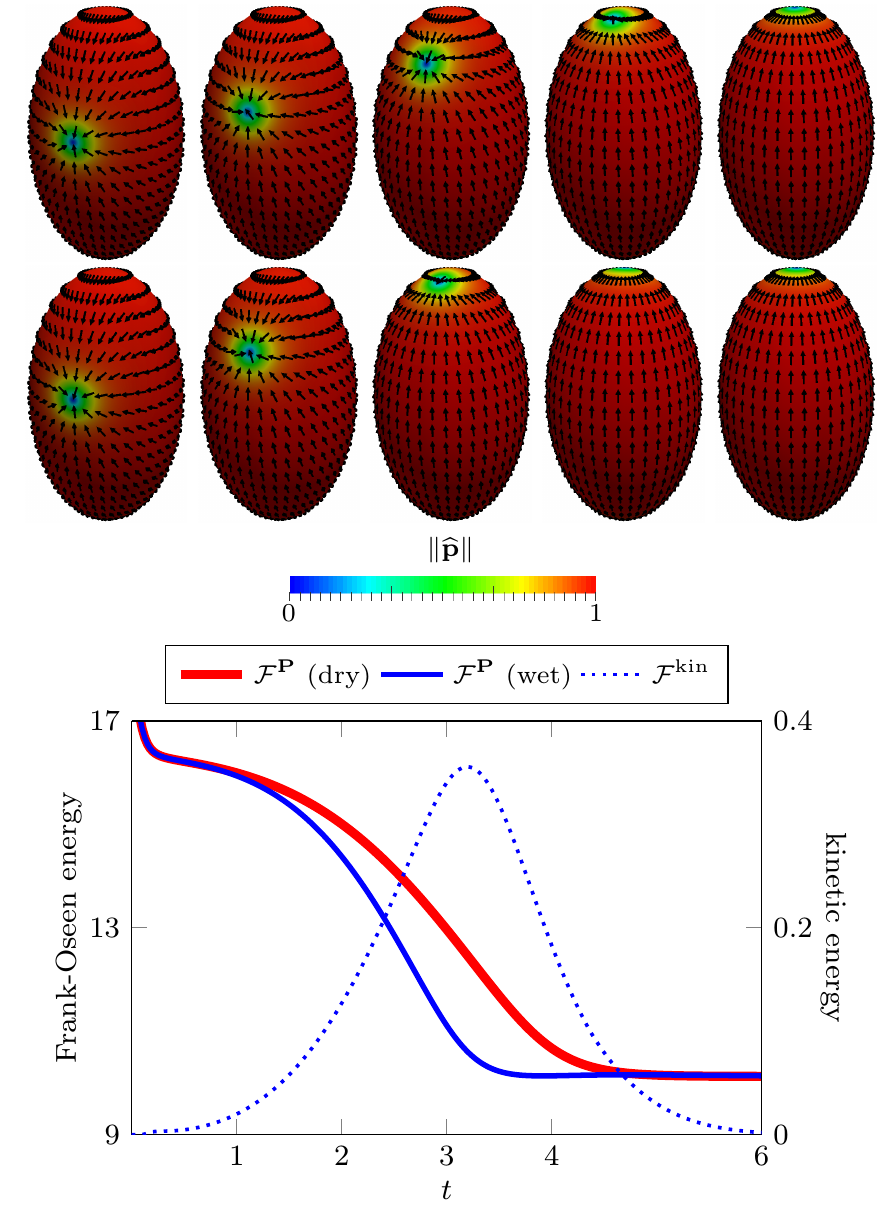}
		\end{minipage}
	}{
		\begin{minipage}{0.49\textwidth}
			\centering
			\def\picHeight{0.28\textwidth}
			\includegraphics[height=\picHeight]{fig_1_dry_solution_0001.png}
			\includegraphics[height=\picHeight]{fig_1_dry_solution_0002.png}
			\includegraphics[height=\picHeight]{fig_1_dry_solution_0003.png}
			\includegraphics[height=\picHeight]{fig_1_dry_solution_0004.png}
			\includegraphics[height=\picHeight]{fig_1_dry_solution_0006.png}\\
			\includegraphics[height=\picHeight]{fig_1_wet_solution_0001.png}
			\includegraphics[height=\picHeight]{fig_1_wet_solution_0002.png}
			\includegraphics[height=\picHeight]{fig_1_wet_solution_0003.png}
			\includegraphics[height=\picHeight]{fig_1_wet_solution_0004.png}
			\includegraphics[height=\picHeight]{fig_1_wet_solution_0006.png} \\
			\insertColorbarHorizontal{ $\|\pExt\|$ }{$0$}{}{}{$1$}
		\end{minipage}
		\begin{minipage}{0.49\textwidth}
			\input{fig_1_energyVsTime.tex}
		\end{minipage}
	}
	\caption{Top: Evolution of the director field $\pExt$ on a stationary ellipsoid of the dry case (top row) and the wet case (bottom row) for $t=1$, $2$, $3$, $4$, $6$ (left to right). Bottom: Surface Frank-Oseen energy $\frankOseenEnergy$ and surface kinetic energy $\kineticEnergy$ vs. time $t$.}
	\label{fig:ellipsoid:results}
\end{figure}

First, we consider eqs. \eqref{eq4} - \eqref{eq6} on a stationary, \ie\ $\normalVelocity=0$, ellipsoidal shape with major axes parameters $(0.7, 0.7, 1.2)$. 
We use the trivial solution as initial condition for the velocity and for the director field $\pExt^0 = \GradSurf\psi^0 / \|\GradSurf\psi^0\|$ with $\psi^0 = x_0/10 + x_1 + x_2/10$ and $\xb = (x_0, x_1, x_2)^T$ the Euclidean coordinate vector. 
The latter generates a vector field with two $+1$ defects -- to be more precise a source and a sink defect -- and an out-of-equilibrium solution. 
Furthermore, we use $\viscosity=2$, $\K=0.6$ and $\meshsize=1.32\cdot10^{-2}$, where $\meshsize$ denotes the maximum mesh size.
\autoref{fig:ellipsoid:results} shows the influence of the hydrodynamics on the dynamical evolution of the director field. The two defects, which fulfill the Poincar\'e-Hopf theorem, 
evolve towards the geometrically favorable positions of high Gaussian curvature, the director field aligns with the minimal curvature lines
of the geometry and as in flat space the hydrodynamics enhances the evolution towards the equilibrium configuration, which coincides for the dry and the wet case.

\begin{figure}[t!]
	\centering
	\ifthenelse{\boolean{submitmode}}{
		\begin{minipage}{0.48\textwidth}
			\centering
			\includegraphics[width=\textwidth]{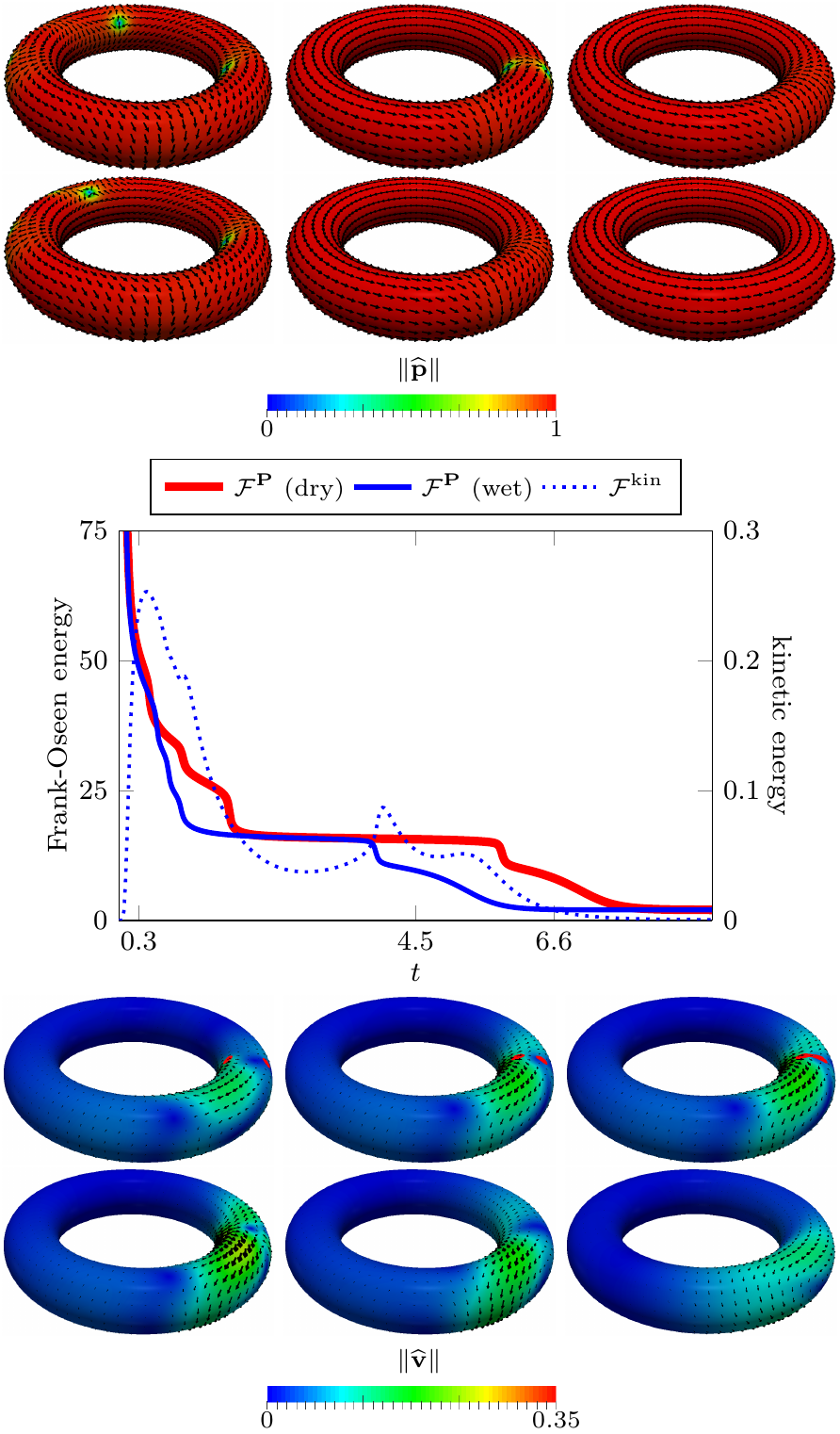}
		\end{minipage}
	}{
		\begin{minipage}{0.49\textwidth}
			\centering
			\def\picWith{0.32\textwidth}
			\includegraphics[width=\picWith]{fig_2_dry_solution_0010.png}
			\includegraphics[width=\picWith]{fig_2_dry_solution_0150.png}
			\includegraphics[width=\picWith]{fig_2_dry_solution_0220.png}
			\\
			\includegraphics[width=\picWith]{fig_2_wet_solution_0010.png}
			\includegraphics[width=\picWith]{fig_2_wet_solution_0150.png}
			\includegraphics[width=\picWith]{fig_2_wet_solution_0220.png}\\
			\insertColorbarHorizontal{ $\|\pExt\|$ }{$0$}{}{}{$1$}
		\end{minipage}
		\begin{minipage}{0.49\textwidth}
			\input{fig_2_energyVsTime.tex}
		\end{minipage}
		\begin{minipage}{0.49\textwidth}
			\def\picWith{0.32\textwidth}
			\includegraphics[width=\picWith]{fig_2_velocity_0100.png}
			\includegraphics[width=\picWith]{fig_2_velocity_0125.png}
			\includegraphics[width=\picWith]{fig_2_velocity_0127.png}
			\includegraphics[width=\picWith]{fig_2_velocity_0135.png}
			\includegraphics[width=\picWith]{fig_2_velocity_0150.png}
			\includegraphics[width=\picWith]{fig_2_velocity_0190.png}
			\insertColorbarHorizontal{ $\|\vExt\|$ }{$0$}{}{}{$0.35$}
		\end{minipage}
	}
	\caption{Top: Evolution of the director field $\pExt$ on a torus of the dry case (top row) and the wet case (bottom row) for $t=0.3$, $4.5$, $6.6$ (left to right). Middle: Surface Frank-Oseen energy $\frankOseenEnergy$ and surface kinetic energy $\kineticEnergy$ vs. time $t$. Bottom: Velocity field $\vExt$ for the annihilation of a source ($+1$, left) and a saddle ($-1$, right) defect in the director field $\pExt$ (red dots) for $t=3$, $3.75$, $3.81$, $4.05$, $4.5$, $5.7$ (left to right and top to bottom).}
	\label{fig:torus:results}
\end{figure}

In the next example we consider a stationary torus with major radius $R=2$, minor radius $r=0.5$ and the $x_2$-axis as symmetry axis. Again we use the trivial solution as initial condition for the velocity $\vExt$ and a random (normalized) vector field for the director field $\pExt$. Here, we use the simulation parameters $\viscosity=1$ and $\K=0.4$. The maximum mesh size is fixed at $\meshsize=2.74\cdot10^{-2}$. All other parameters are equal to that used in \autoref{fig:ellipsoid:results}. In \autoref{fig:torus:results} we focus on the annihilation of defects in one realization. \autoref{fig:torus:results} (top) shows the evolution of the director field $\pExt$ for the dry and the wet case. Again in the wet case the dynamics is enhanced, which is quantified by the stronger overall decay of the surface Frank-Oseen energy, cf. \autoref{fig:torus:results} (middle). Additionally, in \autoref{fig:torus:results} (bottom) the corresponding flow field $\vExt$ is shown for the considered annihilation of a source ($+1$) and a saddle ($-1$) defect in the director field $\pExt$. After all defects are annihilated, which again is in accordance with the Poincar\'e-Hopf theorem, the velocity field $\vExt$ decays to zero and the director field $\pExt$ aligns with the minimal curvature lines of the geometry. The reached equilibrium configuration coincides for both the dry and the wet case. 

\begin{figure}[t!]
	\centering
	\ifthenelse{\boolean{submitmode}}{
		\begin{minipage}{0.48\textwidth}
			\centering
			\includegraphics[width=\textwidth]{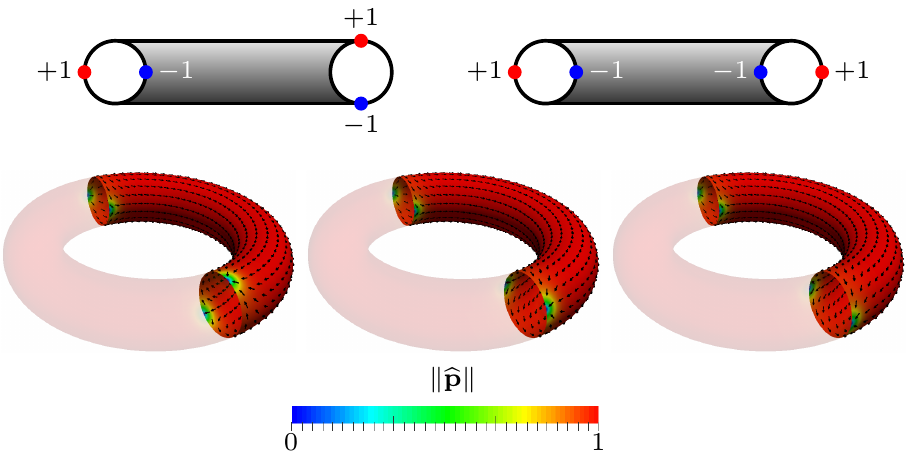}
		\end{minipage}
	}{
		\begin{minipage}{0.5\textwidth}
			\centering
			\begin{tikzpicture}[scale=0.6]
				\def\R{2}			
				\def\r{0.5}			
				\def\defectRadius{0.1}		
				\def\drawOffset{3.5}		
				\def\boundingBoxOffset{1.0}	
				\draw[color=white] (-\R-\r-\drawOffset-\boundingBoxOffset,\r+\boundingBoxOffset) rectangle (\R+\r+\drawOffset+\boundingBoxOffset,-\r-\boundingBoxOffset);
				\shade[top color=black!10,bottom color=black!80] (-\R-\drawOffset,\r) rectangle (\R-\drawOffset,-\r);
				\draw[color=black!100,line width=1.0] (-\R-\drawOffset,\r) -- (\R-\drawOffset,\r) node[] {};
				\draw[color=black!100,line width=1.0] (-\R-\drawOffset,-\r) -- (\R-\drawOffset,-\r) node[] {};
				\draw[color=black!100,line width=1.0,fill=white] (-\R-\drawOffset,0) circle (\r); 
				\draw[color=black!100,line width=1.0,fill=white] (\R-\drawOffset,0) circle (\r); 
				\draw[color=red,fill=red] (-\R-\r-\drawOffset,0) circle (\defectRadius); 
				\draw[color=black] (-\R-\r-\drawOffset,0) node[anchor=east] {\scriptsize$+1$};
				\draw[color=blue,fill=blue] (-\R+\r-\drawOffset,0) circle (\defectRadius); 
				\draw[color=white] (-\R+\r-\drawOffset,0) node[anchor=west] {\scriptsize$-1$};
				\draw[color=red,fill=red] (\R-\drawOffset,\r) circle (\defectRadius); 
				\draw[color=black] (\R-\drawOffset,\r) node[anchor=south] {\scriptsize$+1$};
				\draw[color=blue,fill=blue] (\R-\drawOffset,-\r) circle (\defectRadius); 
				\draw[color=black] (\R-\drawOffset,-\r) node[anchor=north] {\scriptsize$-1$};
				\shade[top color=black!10,bottom color=black!80] (-\R+\drawOffset,\r) rectangle (\R+\drawOffset,-\r);
				\draw[color=black!100,line width=1.0] (-\R+\drawOffset,\r) -- (\R+\drawOffset,\r) node[] {};
				\draw[color=black!100,line width=1.0] (-\R+\drawOffset,-\r) -- (\R+\drawOffset,-\r) node[] {};
				\draw[color=black!100,line width=1.0,fill=white] (-\R+\drawOffset,0) circle (\r); 
				\draw[color=black!100,line width=1.0,fill=white] (\R+\drawOffset,0) circle (\r); 
				\draw[color=red,fill=red] (-\R-\r+\drawOffset,0) circle (\defectRadius); 
				\draw[color=black] (-\R-\r+\drawOffset,0) node[anchor=east] {\scriptsize$+1$};
				\draw[color=blue,fill=blue] (-\R+\r+\drawOffset,0) circle (\defectRadius); 
				\draw[color=white] (-\R+\r+\drawOffset,0) node[anchor=west] {\scriptsize$-1$};
				\draw[color=red,fill=red] (\R+\r+\drawOffset,0) circle (\defectRadius); 
				\draw[color=black] (\R+\r+\drawOffset,0) node[anchor=west] {\scriptsize$+1$};
				\draw[color=blue,fill=blue] (\R-\r+\drawOffset,0) circle (\defectRadius); 
				\draw[color=white] (\R-\r+\drawOffset,0) node[anchor=east] {\scriptsize$-1$};
			\end{tikzpicture}
		\end{minipage} \\
		\begin{minipage}{0.5\textwidth}
			\centering
			\def\picWith{0.32\textwidth}
			\includegraphics[width=\picWith]{fig_3_wet_solution_0001.png}
			\includegraphics[width=\picWith]{fig_3_wet_solution_0005.png}
			\includegraphics[width=\picWith]{fig_3_wet_solution_0025.png}
			\insertColorbarHorizontal{ $\|\pExt\|$ }{$0$}{}{}{$1$}
		\end{minipage}
	}
	\caption{Top: Schematic defect positions of the initial condition (left) and the final configuration (right) on a torus with the analytical initial condition for the director field $\pExt$ and zero initial condition for the velocity field $\vExt$. Red dots are indicating $+1$ defects (sources or sinks) and blue dots are indicating $-1$ defects (saddle). Bottom: Evolution of the director field $\pExt$ for $t=1$, $5$, $25$ (left to right).}
	\label{fig:torus:nontrivialDefectConfig:results}
\end{figure}
\begin{figure}[t!]
	\centering
	\ifthenelse{\boolean{submitmode}}{
		\begin{minipage}{0.48\textwidth}
			\centering
			\includegraphics[width=\textwidth]{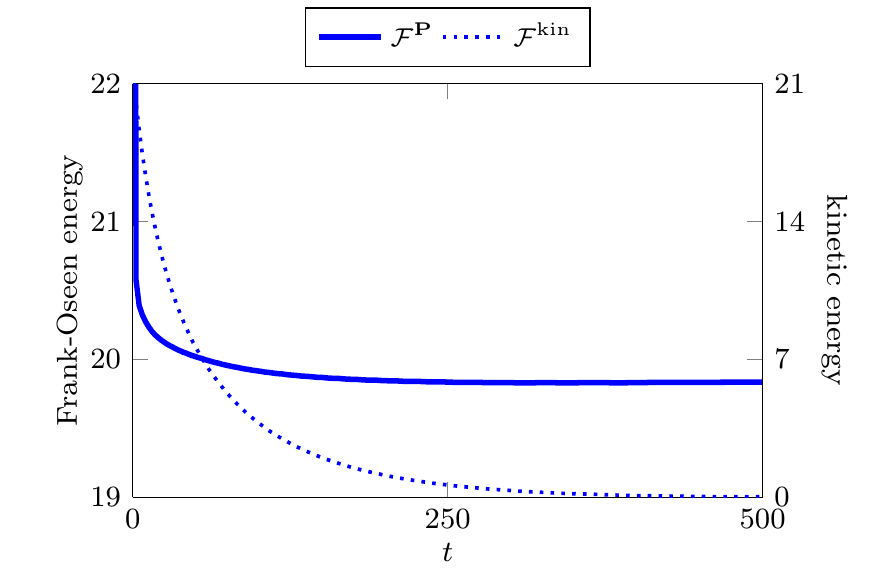}
		\end{minipage}
	}{
		\begin{minipage}{0.49\textwidth}
			\input{fig_4_energyVsTime.tex}
		\end{minipage}
	}
	\caption{Surface Frank-Oseen energy $\frankOseenEnergy$ and surface kinetic energy $\kineticEnergy$ vs. time $t$ for the analytical initial condition for the director field $\pExt$ and the killing vector field as initial condition for the velocity $\vExt$.}
	\label{fig:torus:nontrivialDefectConfig:killingResults}
\end{figure}

While in the two previous examples the expected minimal energy configuration was reached, we now consider an initial condition for which only a local minimum can be reached. 
We use $\pExt^0 = \GradSurf\psi^0 / \|\GradSurf\psi^0\|$ with $\psi^0 = \exp\left(-(\xb-\mb)^2/2\right)$ and $\mb = (R, 0, r)^T$
as initial condition for the director field. 
This produces two $\pm1$ defect pairs which are located in opposite position to each other \wrt\ to the symmetry axis of the torus, again fulfilling the Poincar\'e-Hopf theorem.
Thereby, one pair is rotated by an angle of $\pi/2$ compared to the other along the circle with the small radius, see \autoref{fig:torus:nontrivialDefectConfig:results}. 
The parameters are adapted to $\viscosity=1$, $\K=0.4$ and $\meshsize=2.74\cdot10^{-2}$. In a flat geometry with zero curvature these two pairs would annihilate. However, due to the geometric
interaction in the present case resulting from the difference of the Gaussian curvature inside and outside of the torus, the reached nontrivial defect configuration is stable and the two $\pm1$ defect pairs remain over 
time. The $-1$ defects are attracted to regions with negative Gaussian curvature, \ie\ the inner of the torus, and $+1$ defects are attracted to regions with positive Gaussian curvature, \ie\ the outer of the torus, see \autoref{fig:torus:nontrivialDefectConfig:results}. The reached configuration, is a local minimum with a significantly larger surface Frank-Oseen energy $\frankOseenEnergy$ as the defect-free configuration.
In this example we did not find any significant difference between the dry and the wet case, when the zero initial condition for the velocity $\vExt$ is used.
However, if we use a Killing vector field for the velocity as initial condition, \ie\ $\vExt^0 = 1/2( -x_1 , x_0 , 0 )^T$, \cf \cite{Nitschkeetal_book_2017,Reutheretal_PF_2018}, the four defects start to rotate and cause 
a damping of the flow field, which converges to zero. 
In other words, the defects in the director field produce an additional contribution to the total surface stress tensor and therefore the kinetic energy dissipates to zero, see \autoref{fig:torus:nontrivialDefectConfig:killingResults}. 
The final configuration is a rotation of the configuration reached with $\vExt^0 = 0$, with the rotation angle depending on the strength of the initial velocity and the viscosity.

\begin{figure}[t!]
	\centering
	\ifthenelse{\boolean{submitmode}}{
		\begin{minipage}{0.48\textwidth}
			\centering
			\includegraphics[width=\textwidth]{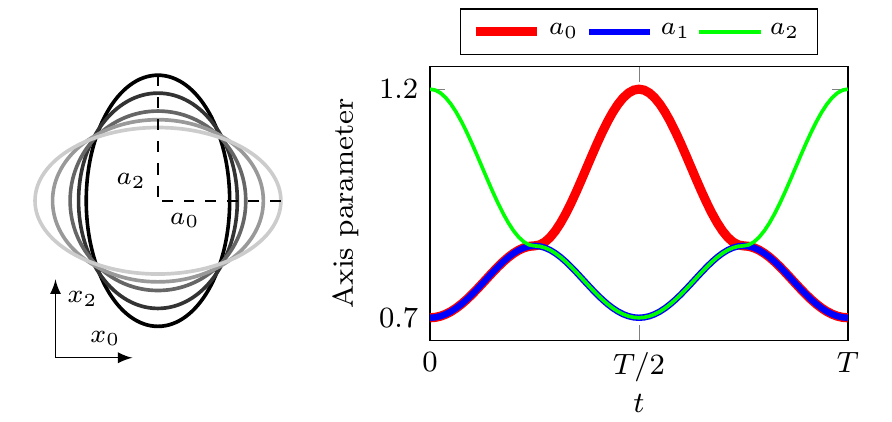}
		\end{minipage}
	}{
		\begin{minipage}{0.5\textwidth}
			\centering
			\begin{minipage}{0.28\textwidth}
				\centering
				\begin{tikzpicture}[scale=1.0]
					\def\rotAngle{0}
					\draw[rotate=\rotAngle,color=black!100,line width=1.0] (0,0) ellipse (0.7000 and 1.2000); 
					\draw[rotate=\rotAngle,color=black!80,line width=1.0] (0,0) ellipse (0.7748 and 1.0287); 
					\draw[rotate=\rotAngle,color=black!60,line width=1.0] (0,0) ellipse (0.8573 and 0.8573); 
					\draw[rotate=\rotAngle,color=black!40,line width=1.0] (0,0) ellipse (1.0287 and 0.7748); 
					\draw[rotate=\rotAngle,color=black!20,line width=1.0] (0,0) ellipse (1.2000 and 0.7000);
					\draw[arrows={-latex}] (-1,-1.5) -- (-0.25,-1.5) node[anchor=south east] {\scriptsize$x_0$};
					\draw[arrows={-latex}] (-1,-1.5) -- (-1,-0.75) node[anchor=north west] {\scriptsize$x_2$};
					\draw[dashed,color=black!100,line width=0.5] (1.2,0) -- (0,0) node[anchor=north west] {\scriptsize$a_0$}; 
					\draw[dashed,color=black!100,line width=0.5] (0,1.2) -- (0,0) node[anchor=south east] {\scriptsize$a_2$};
				\end{tikzpicture}
			\end{minipage}
			\begin{minipage}{0.65\textwidth}
				\centering
				\input{fig_5_ellipsoidalAxes.tex}
			\end{minipage}
		\end{minipage}
	}
	\caption{Left: Schematic description of the ellipsoid evolution for a half period of oscillation. Descending gray scale indicates increasing time. The motion in the second half of the oscillation is reversed, respectively. Right: Major axes parameters for the ellipsoid over a full period of oscillation. The time of one oscillation period is considered to be $T=160$. The major axes parameters are chosen such that the surface area of the ellipsoid is conserved over time.}
	\label{fig:evolvingEllipsoid:schematic}
\end{figure}
\begin{figure}[t!]
	\centering
	\ifthenelse{\boolean{submitmode}}{
		\begin{minipage}{0.48\textwidth}
			\centering
			\includegraphics[width=\textwidth]{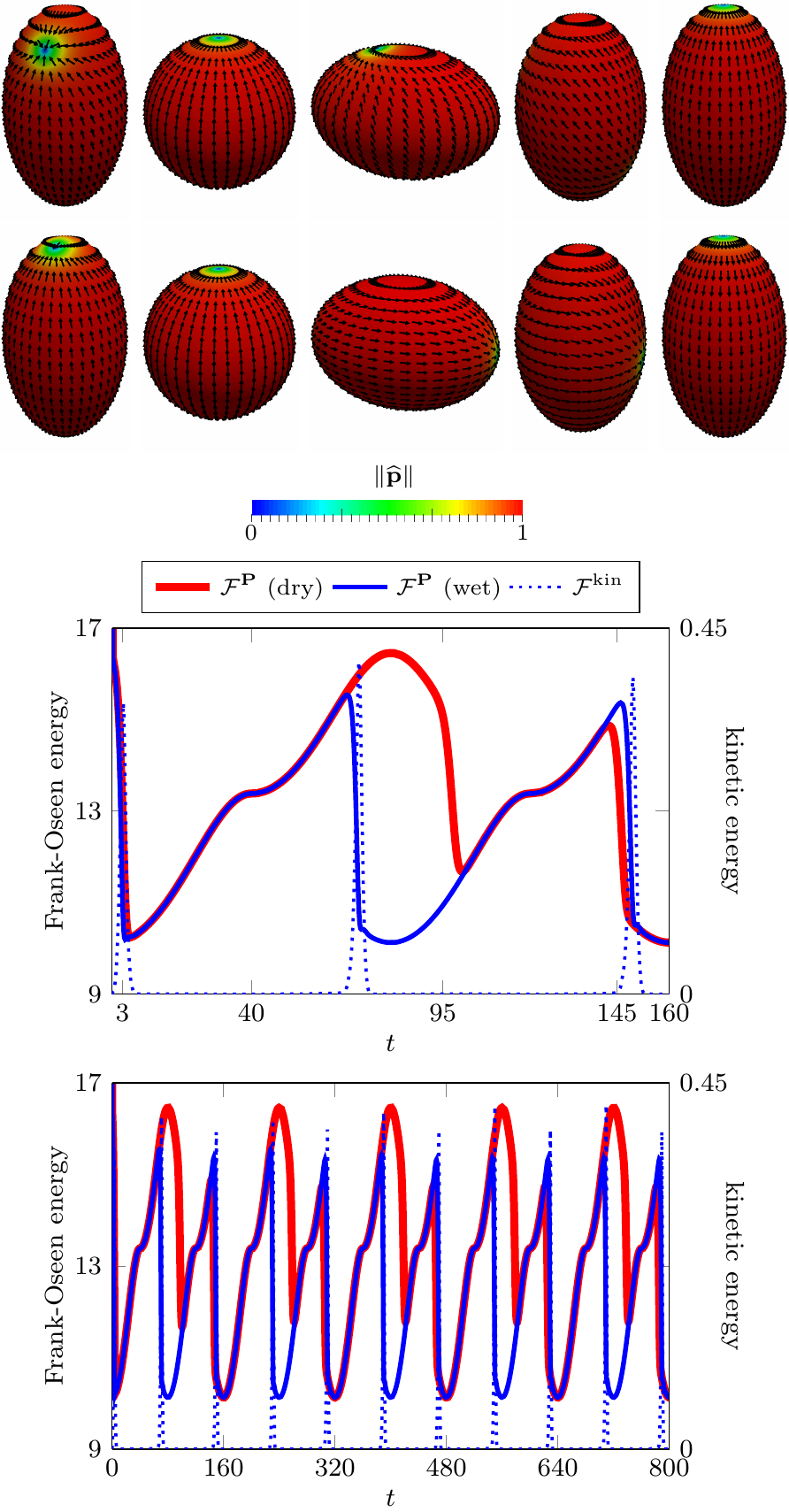}
		\end{minipage}
	}{
		\begin{minipage}{0.49\textwidth}
			\centering
			\def\picHeight{0.28\textwidth}
			\includegraphics[height=\picHeight]{fig_6_dry_solution_0003.png}
			\includegraphics[height=\picHeight]{fig_6_dry_solution_0040.png}
			\includegraphics[height=\picHeight]{fig_6_dry_solution_0095.png}
			\includegraphics[height=\picHeight]{fig_6_dry_solution_0145.png}
			\includegraphics[height=\picHeight]{fig_6_dry_solution_0160.png}\\
			\includegraphics[height=\picHeight]{fig_6_wet_solution_0003.png}
			\includegraphics[height=\picHeight]{fig_6_wet_solution_0040.png}
			\includegraphics[height=\picHeight]{fig_6_wet_solution_0095.png}
			\includegraphics[height=\picHeight]{fig_6_wet_solution_0145.png}
			\includegraphics[height=\picHeight]{fig_6_wet_solution_0160.png}\\
			\insertColorbarHorizontal{ $\|\pExt\|$ }{$0$}{}{}{$1$}
		\end{minipage}
		\begin{minipage}{0.49\textwidth}
			\input{fig_6_energyVsTime_onePeriod.tex}
		\end{minipage}
		\begin{minipage}{0.49\textwidth}
			\input{fig_6_energyVsTime.tex}
		\end{minipage}
	}
	\caption{Top: Evolution of the director field $\pExt$ on an evolving ellipsoid of the dry case (top row) and the wet case (bottom row) for $t=3$, $40$, $95$, $145$, $160$ (left to right). Middle: Surface Frank-Oseen energy $\frankOseenEnergy$ and surface kinetic energy $\kineticEnergy$ vs. time $t$ for the first period of oscillation. Bottom: Surface Frank-Oseen energy $\frankOseenEnergy$ and surface kinetic energy $\kineticEnergy$ vs. time $t$ over five periods of oscillation.}
	\label{fig:evolvingEllipsoid:results}
\end{figure}

We now let the ellipsoid from \autoref{fig:ellipsoid:results} evolve by prescribing the normal velocity $\normalVelocity$, such that the ellipsoid changes to a sphere and afterwards to an ellipsoid with a different axis orientation and vice versa to obtain a shape oscillation. 
The surface area remains constant during the evolution. 
\autoref{fig:evolvingEllipsoid:schematic} shows schematically the evolution of the geometry and the axes parameters for one period of oscillation. 
We use the same simulation parameters and initial conditions as considered in \autoref{fig:ellipsoid:results}. 
The evolution of the director field $\pExt$ is shown in \autoref{fig:evolvingEllipsoid:results}, again for the dry (top) and the wet (bottom) case.  
The defect positions again reallocate at their geometrically favorable position. 
However, due to the change in the geometry the time scale for the reallocation competes with the time-scale for the shape changes. 
The enhanced evolution towards the minimal energy configuration with hydrodynamics becomes even more significant in these situations. 
Already slight modifications of the geometry are enough to push the defect after crossing the sphere configurations (with no preferred defect position) to the energetically favorable state. 
In the dry case there is a strong delay and much stronger shape changes are needed to push the defect to the energetically favorable position. 
First an energy barrier for reallocating the defect position has to be overcome, which is shown by the further increase of the red line after the blue line has already dropped after crossing the sphere configuration in \autoref{fig:evolvingEllipsoid:results} (middle). 
The parameters and the initial condition are further chosen in such a way that the defects in the dry case not quite reach the position at the poles if the shape evolution crosses the sphere.
In the wet case they have moved beyond. 
This results in a constant orientation in the dry case and a flipping of the orientation of the director field in the wet case in each oscillation. 
The final configuration in \autoref{fig:evolvingEllipsoid:results} after completing one oscillation cycle is energetically equivalent for the dry and the wet case even if the orientation of the director field $\pExt$ differs, see also the video in the supplementary material. 
This behavior clearly depends on the used parameters. 
However, it also demonstrates the strong influence hydrodynamics might have in such highly nonlinear systems, where the topology, geometric properties and defect dynamics are strongly coupled. 

These examples together with the demonstrated energy reduction by creation of additional defects in geometrically favored positions in \cite{Nestleretal_JNS_2018}, which is expected to hold also for the wet case, leads to a very rich phase space, considering geometric and material properties, whose exploration is beyond the scope of this paper. 

\section{Discussion}
\label{sec:discussion}
\begin{figure}[t!]
	\centering
	\ifthenelse{\boolean{submitmode}}{
		\begin{minipage}{0.48\textwidth}
			\centering
			\includegraphics[width=\textwidth]{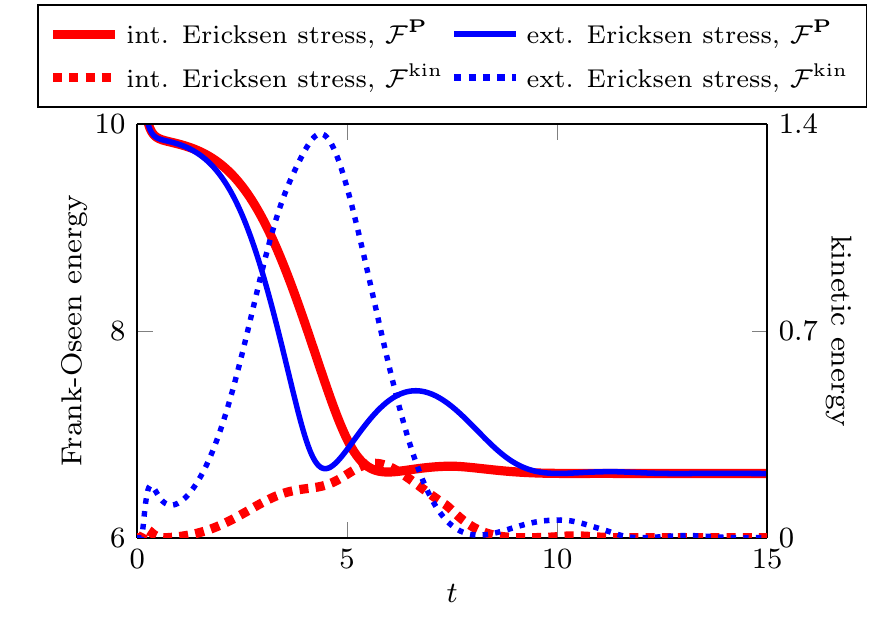}
		\end{minipage}
	}{
		\begin{minipage}{0.49\textwidth}
			\input{fig_7_extIntEricksenStressComparison_energyVsTime.tex}
		\end{minipage}
	}
	\caption{Surface Frank-Oseen energy $\frankOseenEnergy$ and surface kinetic energy $\kineticEnergy$ vs. time $t$ for the simulation with the damped oscillation of the defects around the minimal defect configuration.}
	\label{fig:ellipsoid:Bp_comparison}
\end{figure}

Eqs. \eqref{eq1} - \eqref{eq3} have been derived as a thin-film limit of a three-dimensional simplified Ericksen-Leslie model, see appendix \ref{app:thin_film_limit}. In \cite{Shkoller_CPDE_2002} a similar model was proposed, which differs from eqs. \eqref{eq1} -- \eqref{eq3} with $\normalVelocity=0$ in the extrinsic contributions. Especially the surface Ericksen stress tensor is considered to be $\surfaceEricksenStress = \left(\GradSurf\pb\right)^T\GradSurf\pb$. To show the strong difference between this intrinsic and the extrinsic surface Ericksen stress tensor $\surfaceEricksenStress = \left(\GradSurf\pb\right)^{T} \GradSurf\pb + \left( \shapeOperator\pb \right) \otimes \left( \shapeOperator\pb \right)$ considered here and in \cite{Napolietal_PRE_2016}, we come back to the stationary ellipsoid in \autoref{fig:ellipsoid:results}. 
We use slightly different parameters, \ie\ $\viscosity=0.5$, $\K=0.3$ and $\lambda=1$, which lead to a damped oscillation of the defects around the energetically favorable positions before they reach the final state configuration as in \autoref{fig:ellipsoid:results}. 
In \autoref{fig:ellipsoid:Bp_comparison} the differences in the time evolution of the surface Frank-Oseen energy as well as the surface kinetic energy are shown for both cases the intrinsic and extrinsic surface Ericksen stress tensor. 
The influence of the hydrodynamics is much stronger for the extrinsic surface Ericksen stress. Together with the example in \autoref{fig:evolvingEllipsoid:results} such differences in the dynamics might have a huge impact on the overall evolution if also shape changes are considered. 

\begin{figure}[t!]
	\centering
	\ifthenelse{\boolean{submitmode}}{
		\begin{minipage}{0.48\textwidth}
			\centering
			\includegraphics[width=\textwidth]{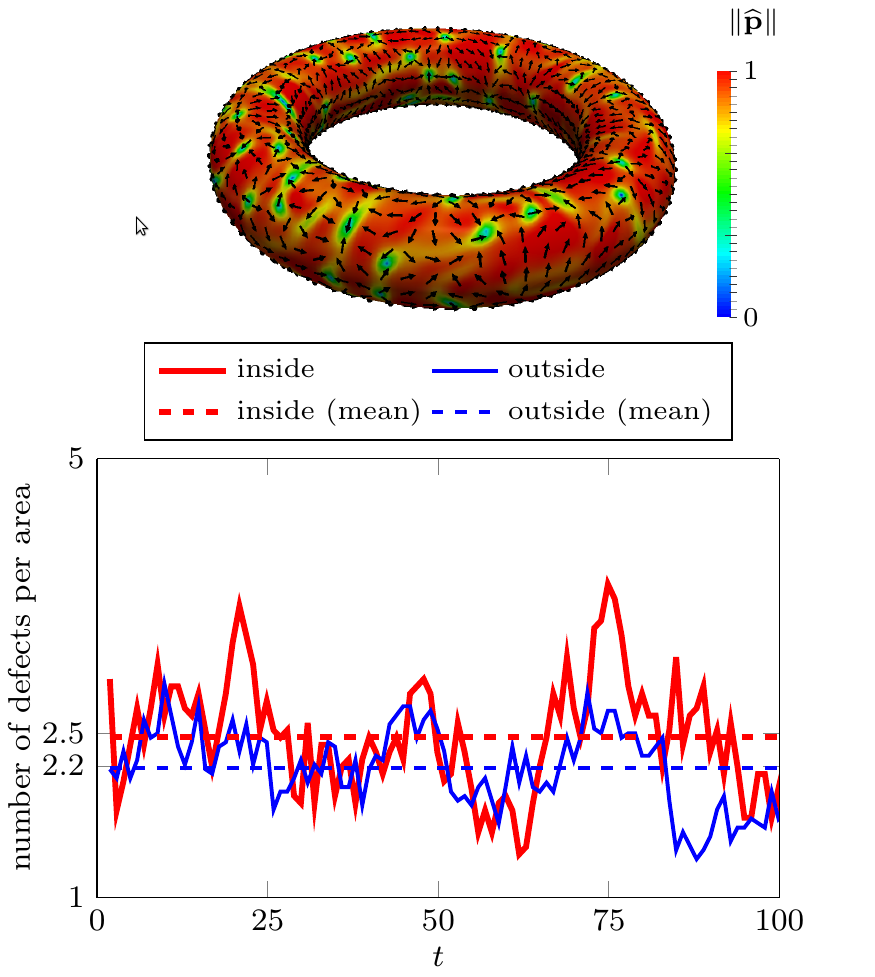}
		\end{minipage}
	}{
		\begin{minipage}{0.49\textwidth}
			\centering
			\begin{minipage}{0.2\textwidth}
				\centering
				\phantom{\insertColorbarVertical{ $\|\pExt\|$ }{$0$}{}{}{$1$}}
			\end{minipage}
			\begin{minipage}{0.49\textwidth}
				\centering
				\def\picwidth{\textwidth}
				\includegraphics[width=\picwidth]{fig_8_active_solution_0145.png}
			\end{minipage}
			\begin{minipage}{0.2\textwidth}
				\centering
				\insertColorbarVertical{ $\|\pExt\|$ }{$0$}{}{}{$1$}
			\end{minipage}
		\end{minipage}
		\begin{minipage}{0.49\textwidth}
			\input{fig_8_defectsVsTime.tex}
		\end{minipage}
	}
	\caption{Top: Snapshot of the director field $\pExt$ of the surface active polar viscous gel model on the torus for $t = 14.3$. Bottom: Number of defects per area for the inner ($\gaussianCurvature > 0$) and outer ($\gaussianCurvature < 0$) region of the torus vs. time $t$.}
	\label{fig:active_polar_gel}
\end{figure}

All results so far are for the simplified surface Ericksen-Leslie model. However,  appendix \ref{app:thin_film_limit} provides all necessary tools to do the thin-film analysis also for more complicated systems, such as more general Ericksen-Leslie models or active versions of them.
Here, we provide the formulation for a surface active polar viscous gel, see \cite{Simhaetal_PRL_2002,Kruseetal_PRL_2004} and \cite{Tjhungetal_PNAS_2012,Marthetal_JRSI_2015} for the considered three-dimensional formulation, which correspond to eqs. \eqref{eqNSTF} - \eqref{eqDirTF} with boundary conditions \eqref{eqNormalPBC} - \eqref{eqNavierBC}, \ie\
\begin{align*}
	\partial_{t}\veloTF + \dirnablaTF{\relveloTF}\veloTF &= -\nablaTF\pressTF + \viscosity\vecLaplace\veloTF + \divTF\activeStress - \lambda\divTF\ericksenStress \\
	\divTF\veloTF &= 0 \\
	\partial_{t}\pTF + \dirnablaTF{\relveloTF}\pTF &= \hTF + \alpha\DTF\pTF + \OmegaTF\pTF
\end{align*}
with  
\begin{align*}
	\activeStress &= \frac{1}{2}\left(\pTF\otimes\hTF - \hTF\otimes\pTF\right) - \frac{\alpha}{2}\left(\pTF\otimes\hTF + \hTF\otimes\pTF\right) \\
	&\qquad+ \beta\pTF\otimes\pTF\\
	\hTF &= \K\vecLaplace\pTF - \Kn\left( \left\| \pTF \right\|_{\TFilm}^{2} - 1 \right)\pTF \\
	\DTF &= \frac{1}{2}\left(\Grad\veloTF + \left(\Grad\veloTF\right)^T\right) \\
	\OmegaTF &= \frac{1}{2}\left(\Grad\veloTF - \left(\Grad\veloTF\right)^T\right)
\end{align*}
and $\alpha,\beta\in\R$. 
The Navier-Stokes equation now contains additional distortion and active stresses, combined in $\activeStress$, while in the director field equation additional contributions from the strain rate tensor $\DTF$ and the vorticity tensor $\OmegaTF$ arise. The corresponding thin-film limit reads
\begin{align}
	\label{eq:active:surface:1}
	\ProjectSurf\partial_{t}\vb + \dirnablaS{\vb}\vb &=  \normalVelocity\left( \shapeOperator\vb + \GradSurf\normalVelocity \right) + \viscosity\left( -\laplaceDeRham\vb + 2\gaussianCurvature\vb \right) \notag \\
	&\quad + \viscosity\left( \GradSurf\left(\normalVelocity\meanCurvature\right) - 2\DivSurf\left(\normalVelocity\shapeOperator\right) \right) \notag \\
	&\quad -\GradSurf\pressure + \DivSurf\surfaceActiveStress - \lambda\DivSurf\surfaceEricksenStress \notag \\
	&\quad -\frac{1-\alpha}{2}(\pb^T\shapeOperator(\meanCurvature\vb + \GradSurf\normalVelocity))\pb \notag \\
	&\quad + \frac{1+\alpha}{2}(\pb^T\shapeOperator\vb)\shapeOperator\pb \\
	\label{eq:active:surface:2}
	\DivSurf\vb &= \normalVelocity\meanCurvature\\
	\label{eq:active:surface:3}
	\ProjectSurf\partial_{t}\pb + \dirnablaS{\vb}\pb &= \hb + \alpha\DS\pb + \OmegaS\pb - \alpha\normalVelocity\shapeOperator\pb
\end{align}
with
\begin{align*}
	\surfaceActiveStress &= \frac{1}{2}\left(\pb\otimes\hb - \hb\otimes\pb\right) - \frac{\alpha}{2}\left(\pb\otimes\hb + \hb\otimes\pb\right) \\
	&\qquad+ \beta\pb\otimes\pb \\
	\hb &= \K\left( \laplaceDivGrad\pb - \shapeOperator^{2}\pb \right) -\Kn\left( \left\| \pb \right\|_{\surf}^{2} - 1 \right)\pb \\
	\DS &= \frac{1}{2}\left(\GradSurf\vb + \left(\GradSurf\vb\right)^T\right) \\
	\OmegaS &= \frac{1}{2}\left(\GradSurf\vb - \left(\GradSurf\vb\right)^T\right).
\end{align*}
Besides the corresponding surface operators and the additional geometric coupling terms with the shape operator $\shapeOperator$ and the mean curvature $\meanCurvature$ we also obtain an explicit appearance of the normal velocity $\normalVelocity$ in the director field equation. Overall the additional terms in the more general Ericksen-Leslie model lead to an even tighter coupling between geometric properties and dynamics. 
The described numerical approach, see appendix \ref{app:numerics}, can be adapted to also solve the surface active polar viscous gel model proposed in eqs. \eqref{eq:active:surface:1} - \eqref{eq:active:surface:3}.  \autoref{fig:active_polar_gel} shows results on a torus.  
We use the same torus and the same initial conditions for the velocity field $\vExt$ and the director field $\pExt$ as considered in \autoref{fig:torus:results}. 
The parameters are adapted to $\alpha = 1.1$, $\beta = 20$, $\lambda = 0.02$, $\viscosity = 200$, $\K = 0.01$ and $\Kn = 5$, while all other parameters remain unchanged.
\autoref{fig:active_polar_gel} (top) shows a snapshot of the director field, see also the video in the supplementary material. 
The significantly reduced parameter $\eta$ yields smeared out defects and promotes the annihilation and creation of new defects. 
In \autoref{fig:active_polar_gel} (bottom) the number of defects per area against the time is plotted. 
Thereby, we distinguish between the inner ($\gaussianCurvature<0$) and the outer ($\gaussianCurvature>0$) region of the torus and observe slightly more defects per area in the inner part. 
This might be due to the stronger geometric force (resulting from a higher absolute value of the Gaussian curvature in the inner region), the continuous creation and annihilation of defects as well as the fact that defects of opposite topological charge are attracted to each other.
As in the passive case, a detailed analysis of such phenomena has to be discussed elsewhere. 

\begin{acknowledgments}
	This work was financially supported by the German Research Foundation (DFG) through project Vo899-19. We used computing resources provided by Jülich Supercomputing Centre within project HDR06.
\end{acknowledgments}

\appendix

\section{Numerics}
\label{app:numerics}

To efficiently solve the surface Navier-Stokes equation in \cite{Reutheretal_PF_2018} a heavy assembly workload was avoided by applying $\surfNormal\times$ to eq. \eqref{eq4} and considering the rotated velocity field $\wExt := \surfNormal\times\vExt$. 
Since only the $\BigRotSurf\RotSurf\left(\cdot\right)$ operator occurs as second order operator we here can also use the same idea to reduce the assembly costs. 
Thus, the rotated version of eqs. \eqref{eq4} and \eqref{eq5} with tangential penalization of the rotated velocity $\wExt$ now reads 
\begin{align}
	\label{eq4.rotated}
	\ProjectSurf\partial_{t}\wExt + \nabla_{\wExt} \wExt &= - \RotSurf p + \viscosity \left(\GradSurf\DivSurf\wExt + 2 \gaussianCurvature \wExt \right) \nonumber \\
	&\qquad + \mathbf{f}_g + \mathbf{f}_{\wExt} - \lambda \surfNormal\times \DivSurf\surfaceEricksenStressTilde \\
	\label{eq5.rotated}
	\RotSurf\wExt &= \normalVelocity \meanCurvature
\end{align}
where we used for convenience the abbreviations 
\begin{align*}
	\mathbf{f}_g &:= \normalVelocity \RotSurf \normalVelocity + 2\viscosity \left( \meanCurvature\RotSurf \normalVelocity - \surfNormal\times\left(\shapeOperator\GradSurf \normalVelocity\right) \right)\\
	\mathbf{f}_{\wExt} &:= - \normalVelocity \surfNormal\times\left(\shapeOperator \left(\surfNormal\times\wExt\right)\right) - \tanPenV(\wExt\cdot\surfNormal)\surfNormal
\end{align*}
and the alternative form of the viscous terms proposed in \eqref{eq:viscouse_alternative}.

\subsection*{Time discretization} 
For the discretization in time we again use the same approach proposed in \cite{Reutheretal_PF_2018}. Let the time interval $[0,t_{\textup{end}}]$ be divided into a sequence of discrete times $0 < t^0 < t^1 < ...$ with time step width $\tau^m = t^{m} - t^{m-1}$. Thereby, the superscript denotes the timestep number. The vector field $\wExt^m(\xb)$ correspond to the respective rotated velocity field $\wExt(\xb, t^m)$. All other quantities follow the same notation. The time derivative is approximated by a standard difference quotient and a Chorin projection method \cite{Chorin_MC_1968} is applied to eqs. \eqref{eq4.rotated} and \eqref{eq5.rotated}. Furthermore, we define the discrete time derivatives $\dTau{\wExt} := \frac{1}{\tau^m}\left( \wExt^* - \ProjectSurf\wExt^{m-1} \right)$ and $\dTau{\pExt} := \frac{1}{\tau^m}\left( \pExt^{m} - \ProjectSurf\pExt^{m-1} \right)$, with $ \ProjectSurf$ the projection to the surface at time $t^{m}$. Thus, we get a time-discrete version of eqs. \eqref{eq4.rotated}, \eqref{eq5.rotated} and \eqref{eq6}
\begin{align}
	\label{eq4.rotated.timediscrete}
	\dTau{\wExt} + \nabla_{\wExt^{m-1}} \wExt^* &= \viscosity \left(\GradSurf\DivSurf\wExt^* + 2 \gaussianCurvature \wExt^* \right) + \mathbf{f}_g + \mathbf{f}_{\wExt^*} \notag \\
	&\hspace*{0.5cm} - \lambda \surfNormal\times \DivSurf\surfaceEricksenStressTilde  \\
	\label{eq5.rotated.timediscrete}
	\tau^m\laplaceBeltrami p^{m} &= \RotSurf\wExt^* - \normalVelocity \meanCurvature \\
	\label{pressureCorrection}
	\wExt^{m} &= \wExt^* - \tau^m\RotSurf p^{m} \\
	\label{eq6.timediscrete}
        \dTau{\pExt} + \nabla_{\vExt^{m}}\pExt^{m} &= \K \left( \laplaceDivGradTilde\pExt^{m} - \shapeOperator^2 \pExt^{m} \right) \notag \\
        &\hspace*{0.5cm} - \Kn \left( \| \pExt^{m-1} \|^2 - 1 \right) \pExt^{m} \notag \\
        &\hspace*{0.5cm} - \tanPenP \left(\surfNormal  \cdot \pExt^{m} \right) \surfNormal 
\end{align}
where $\surfaceEricksenStressTilde$ is evaluated at the old timestep, \ie\ $\surfaceEricksenStressTilde = \left(\GradSurf\pExt^{m-1}\right)^{T} \GradSurf\pExt^{m-1} + \left( \shapeOperator\pExt^{m-1} \right) \otimes \left( \shapeOperator\pExt^{m-1} \right)$.
Note that for readability we used a Taylor-$0$ linearization of the transport term in \eqref{eq4.rotated.timediscrete} and the norm-$1$ penalization term in \eqref{eq6}. In the simulations from above we performed a Taylor-$1$ linearization, see \cite{Nitschkeetal_book_2017} and \cite{Nestleretal_JNS_2018} for details.
 
\subsection*{Spatial discretization}
The considered extension of the tangential vector fields to the Euclidean space allows us to apply the surface finite element method \cite{Dziuketal_AN_2013} for each component of the respective vector field. Let $\surfDiscrete$ denote the interpolation of the surface $\surf(t^m)$ at time $t=t^m$ such that $\surfDiscrete := \bigcup_{T \in \triangulation} T$ with a conforming triangulation $\triangulation$. Furthermore, we introduce the finite element space 
\begin{equation*}
	\mathbb{V}_h(\surfDiscrete) = \left\{ v_h \in C^0(\surfDiscrete) \,:\, v_h|_T \in \mathbb{P}^1, \, \forall \, T \in \triangulation \right\}
\end{equation*}
which is used twice as trail and test space and the standard $L_2$ scalar product on $\mathbb{V}_h(\surfDiscrete)$,
$(\alpha,\beta) := \int_\surfDiscrete \left\langle\alpha,\beta\right\rangle \dS \formPeriod$
By using an operator splitting technique we decouple the hydrodynamic and the director field equation in the following way. First the surface finite element approximation of eqs. \eqref{eq4.rotated.timediscrete}, \eqref{eq5.rotated.timediscrete} is solved, which reads: find $\wExt^*_i,p^{m}\in \mathbb{V}_h(\surfDiscrete)$ s.t. $\forall\xi,\eta\in\mathbb{V}_h(\surfDiscrete)$
\begin{align*}
	&\weak{(\dTau{\wExt})_i + \nabla_{\wExt^{m-1}} \wExt_i^* - 2\viscosity\gaussianCurvature\wExt_i^* - (\mathbf{f}_g + \mathbf{f}_{\wExt^*})_i }{\xi}  \nonumber \\
	&\quad= - \weak{ \viscosity\DivSurf\wExt^*}{(\GradSurf\xi)_i}  - \weak{ \lambda\surfaceEricksenStressTilde }{ \GradSurf\left(\surfNormal\times(\xi\eb_i)\right) } \\
	&\weak{ \tau^m\GradSurf p^{m} }{ \GradSurf\eta } = \weak{ \normalVelocity \meanCurvature - \RotSurf\wExt^* }{ \eta }
\end{align*}
for $i=x,y,z$. The resulting vector field $\wExt^*$ is then used to determine $\wExt^{m}$ by the pressure correction step in eq. \eqref{pressureCorrection}. The transformation $\vExt^{m} = -\surfNormal\times\wExt^{m}$ leads to the velocity field at the new timestep $t^{m}$. Finally, the surface finite element approximation of eq. \eqref{eq6.timediscrete} is solved, which reads: find $\pExt^{m}_i\in \mathbb{V}_h(\surfDiscrete)$ s.t. $\forall\xi\in\mathbb{V}_h(\surfDiscrete)$
\begin{align*}
	&\weak{ (\dTau{\pExt})_i + \nabla_{\vExt^{m}}\pExt^{m}_i }{ \xi } \nonumber \\
	&\quad= \weak{ \K\GradSurf\pExt^{m} }{ \GradSurf(\xi\eb_i) } - \weak{ \K\shapeOperator^2\pExt^{m}_i }{ \xi } \nonumber \\
	&\qquad- \weak{ \Kn \left( \| \pExt^{m-1} \|^2 - 1 \right) \pExt^{m}_i + \tanPenP \left(\surfNormal  \cdot \pExt^{m} \right) \surfNormal_i }{ \xi }
\end{align*}
for $i=x,y,z$.

\subsection*{Pressure relaxation schemes}
In some situations it is useful to modify the Chorin projection scheme \eqref{eq4.rotated.timediscrete}, \eqref{eq5.rotated.timediscrete} and \eqref{pressureCorrection}. To be more precise the resulting finite element matrix of the pressure equation \eqref{eq5.rotated.timediscrete} is sometimes ill-conditioned, especially when the term $\RotSurf\wExt^*$ is big compared to the others. 
The solution of eq. \eqref{eq5.rotated.timediscrete} can be seen as the steady-state solution of a heat conduction equation where the heat source is determined by the right hand side of eq. \eqref{eq5.rotated.timediscrete}. 
Therefore, we add a relaxation scheme in form of a discrete time derivative on a different timescale to the left hand side of eq. \eqref{eq5.rotated.timediscrete}, \ie 
\begin{align}
	\label{eq5.rotated.timediscrete.relaxed}
	&\frac{1}{\tau^*}\left( p^{m+1,l+1} - p^{m+1,l} \right) - \tau^m\laplaceBeltrami p^{m+1,l+1} \nonumber \\
	&\qquad= -\RotSurf\wExt^* + \normalVelocity \meanCurvature \formComma
\end{align}
where $\tau^*$ denotes the timestep and $l$ the timestep number on the different timescale. Instead of solving eq. \eqref{eq5.rotated.timediscrete} the iterative process in eq. \eqref{eq5.rotated.timediscrete.relaxed} is performed until a steady-state is reached, which is then used in the correction step in eq. \eqref{pressureCorrection}.

\section{Thin film limit}
\label{app:thin_film_limit}
We assume a regular moving surface $\surf(t)\subset\R^{3}$ without boundaries and a thin film $\TFilm(t):=\surf(t)\times[-h/2,h/2]\subset\R^{3}$ of sufficiently small thickness $h$, 
such that the thin film parametrization $ \paraTF(t,\ucoord,\vcoord,\ncoord) = \paraS(t,\ucoord,\vcoord) + \ncoord\surfNormal(t,\ucoord,\vcoord) $ is injective for
the surface parametrization $ \paraS(t,\cdot,\cdot) $. Thereby, $ \ucoord$ and $\vcoord $ denote the local surface coordinates, 
$ \surfNormal(t,\cdot,\cdot) $ the surface normal field and $\ncoord\in[-h/2,h/2]$ is the local normal coordinate.
Since the thin film is moving according to the surface, the parametrization $ \paraTF $ is not unique, which arises from the choice of an observer within the thin film.
For a pure Eulerian observer, \ie\ for the observer velocity $ \obveloTF = \partial_{t}\paraTF = 0 $, we are not able to formulate proper intrinsic physics at the surface $ \surf $ for $ h\rightarrow 0 $.
To overcome this issue, we choose a transversal observer as the surface observer parametrization $ \paraS $, \ie\ Eulerian in the tangential space and Lagrangian in normal direction and
hence $ \partial_{t}\paraS = \partial_{t}\paraTF\atSurf = \normalVelocity\surfNormal $, where $ \normalVelocity = \obveloTFC^{\ncoord}\atSurf$ is the normal surface velocity of $ \surf $.
To ensure a constant thickness $ h $ of the thin film and that the surface $\surf$ is located in the middle of the thin film over time,
we stipulating the same transversal behavior for both boundary surfaces. 
Thus, we get $\partial_{t}\paraTF\big|_{\partial\TFilm} = \normalVelocity\surfNormal \pm \frac{h}{2}\partial_{t}\surfNormal =\obveloTFC^{\ncoord}\big|_{\partial\TFilm}\surfNormal$ and therefore $\obveloTFC^{\ncoord}\big|_{\partial\TFilm} = \normalVelocity $, since $ \partial_{t}\surfNormal $ is always tangential on the boundaries $\partial\TFilm$.

For notational compactness of tensor algebra we use the thin film calculus presented in \cite{Nitschkeetal_PRSA_2018} and \cite{Nestleretal_JNS_2018} (Appendix) which is based on Ricci calculus, 
where lowercase indices $i,j,k,\ldots$ 
denote components \wrt\ $ \ucoord$ and $\vcoord$ in the surface coordinate system and uppercase indices $I,J,K,\ldots$ denote
components \wrt\ $ \ucoord, \vcoord  $ and $ \ncoord $ in the extended three dimensional thin film coordinate system.
Metric quantities at the surface $\surf$ are the metric tensor $ \gSC_{ij} = \left\langle \partial_{i}\paraS, \partial_{j}\paraS \right\rangle_{\R^{3}} $ (first fundamental form),
the shape operator $ \shapeOperatorc_{ij} = -\left\langle \partial_{i}\paraS, \partial_{j}\surfNormal \right\rangle_{\R^{3}} $ (second fundamental form),
its square $ \left[ \shapeOperator^{2} \right]_{ij} = \shapeOperatorc_{ik}\tensor{\shapeOperatorc}{^{k}_{j}} = \left\langle \partial_{i}\surfNormal, \partial_{j}\surfNormal \right\rangle_{\R^{3}} $ (third fundamental form),
the mean curvature $ \meanCurvature = \tr\shapeOperator = \tensor{\shapeOperatorc}{^{i}_{i}} $, the Gaussian curvature $ \gaussianCurvature = \det\shapeOperator^{\sharp} = \det\left\{ \shapeOperatorc^{ij} \right\} $ and
the Christoffel symbols $ \GammaS_{ij}^{k}=\frac{1}{2}\gSC^{kl}(\partial_i \gSC_{jl} + \partial_j \gSC_{il} - \partial_l \gSC_{ij}) $ for covariant differentiating (\eg\ 
$ \tensor{\left[ \GradSurf\pb \right]}{^{i}_{k}} = \tensor{\pbc}{^{i}_{|k}} = \partial_{k}\pbc^{i} + \GammaS_{kj}^{i}\pbc^{j}$ 
for a contravariant vector field $ \pb\in\Tangent\surf = \Tangent^{1}\surf $).
In the thin film $ \TFilm $, the metric tensor $ \gTFC_{IJ} = \left\langle \partial_{I}\paraTF, \partial_{J}\paraTF \right\rangle_{\R^{3}} $ and the Christoffel symbols
$ \GammaTF_{IJ}^{K}=\frac{1}{2}\gTFC^{KL}(\partial_I \gTFC_{JL} + \partial_J \gTFC_{IL} - \partial_L \gTFC_{IJ}) $ for covariant differentiating (\eg\ 
$ \tensor{\left[ \nablaTF\pTF \right]}{^{I}_{K}} = \tensor{\pTFC}{^{I}_{;K}} = \partial_{K}\pTFC^{I} + \GammaTF_{KJ}^{I}\pTFC^{J}$  
for a contravariant vector field $ \pTF\in\Tangent\TFilm = \Tangent^{1}\TFilm $) can be developed at the surface by
$ \gTFC_{ij} = \gSC_{ij} -2\ncoord\shapeOperatorc_{ij} + \ncoord^{2}\left[ \shapeOperator^{2} \right]_{ij} $, $ \gTFC_{\ncoord\ncoord} = 1 $, $ \gTFC_{\ncoord i} = \gTFC_{i \ncoord} = 0 $,
$ \gTFC^{ij} = \gSC^{ij} + \landauNor $, $ \gTFC^{\ncoord\ncoord} = 1 $, $ \gTFC^{\ncoord i} = \gTFC^{i \ncoord} = 0 $,
$\GammaTF_{ij}^{k} = \GammaS_{ij}^{k} + \landauNor  $, $\GammaTF_{ij}^{\ncoord} = \shapeOperatorc_{ij} + \landauNor  $,
$\GammaTF_{i\ncoord}^{k} = \GammaTF_{\ncoord i}^{k} = -\tensor{\shapeOperatorc}{^{k}_{i}} + \landauNor  $ and 
$\GammaTF_{I\ncoord}^{\ncoord} =  \GammaTF_{\ncoord I}^{\ncoord} = \GammaTF_{\ncoord \ncoord}^{K} = 0$, see \cite{Nitschkeetal_PRSA_2018} and \cite{Nestleretal_JNS_2018} for
details. 

Our starting point is the simplified local three dimensional Ericksen-Leslie model \cite{Linetal_ARMA_2000}, \ie
\begin{align}
	\partial_{t}\veloTF + \dirnablaTF{\relveloTF}\veloTF &= -\nablaTF\pressTF + \viscosity\vecLaplace\veloTF - \lambda\divTF\ericksenStress \label{eqNSTF}\\
	\divTF\veloTF &= 0 \label{eqContiTF}\\
	\partial_{t}\pTF + \dirnablaTF{\relveloTF}\pTF &= \K\vecLaplace\pTF - \Kn\left( \left\| \pTF \right\|_{\TFilm}^{2} - 1 \right)\pTF \label{eqDirTF}
\end{align}
in $ \TFilm\times\R_{+} $ with fluid velocity $ \veloTF\in\Tangent\TFilm $, relative fluid velocity $ \relveloTF = \veloTF - \obveloTF\in\Tangent\TFilm $, with respect to the observer velocity 
$ \obveloTF = \partial_{t}\paraTF$, director field $ \pTF\in\Tangent\TFilm $, pressure $ \pressTF $, Ericksen stress tensor $\ericksenStress = \left(\nablaTF\pTF\right)^{T} \nablaTF\pTF$, fluid viscosity $\viscosity$, competition between kinetic and elastic potential energy $\lambda$ and elastic relaxation time for the molecular orientation field $\K$. 
Besides initial conditions, we consider homogeneous Dirichlet boundary conditions for the normal components and Neumann boundary conditions for the tangential components of the director 
and homogeneous Navier boundary conditions for the velocity field, \ie
\begin{align}
  \left\langle \pTF , \surfNormal \right\rangle_{\TFilm} &= \pTFC_{\ncoord} = 0\label{eqNormalPBC}\\
  \dirnablaTF{\surfNormal}\left(\ProjectPartialTF\pTF\right)^{\flat}
      &= \left\{ \pTFC_{i;\ncoord} \right\} = 0 \label{eqNeumannBC}\\      
  \ProjectPartialTF\left( \surfNormal\cdot\Lie{\veloTF}\gTF  \right) 
      &= \left\{ \veloTFC_{i;\ncoord} + \veloTFC_{\ncoord;i} \right\} = 0 \label{eqNavierBC}
\end{align}
at the boundaries $ \partial\TFilm $ in its covariant form. Thereby, $ \Lie{\veloTF}\gTF $ denotes the viscous stress tensor in terms of the Lie derivative $ \Lie{} $ and 
$\ProjectPartialTF: \Tangent_{1}\TFilm\big|_{\partial\TFilm} \rightarrow\Tangent_{1}\partial\TFilm $ is the orthogonal covariant projection into the covariant boundary tangential space.
Note that it holds
\begin{align*}
  \left\langle \veloTF , \surfNormal \right\rangle_{\TFilm} &= \veloTFC_{\ncoord} = \obveloTFC_{\ncoord} = \normalVelocity
\end{align*}
on $\partial\TFilm$, which follows from the special choice of the transversal observer from above.

As proposed in \cite{Nitschkeetal_PRSA_2018} the boundary quantities are continuable to the surface $\surf$ by Taylor expansions at the boundaries, which,
\eg, results in 
\begin{align}
  \pTFC_{\ncoord}\atSurf &= \landauh \label{eqNormalPBCAtS}
 &\pTFC_{\ncoord;\ncoord}\atSurf &= \landauh \\
  \pTFC_{i;\ncoord}\atSurf &= \landauh
 &\pTFC_{i;\ncoord;\ncoord}\atSurf &= \landauh\\
  \veloTFC_{\ncoord;\ncoord}\atSurf &= \landauh
 &\veloTFC_{\ncoord;\ncoord;\ncoord}\atSurf &= \landauh \label{eqNormalVeloBCAtS}\\
  \veloTFC_{i;\ncoord}\atSurf + \veloTFC_{\ncoord;i}\atSurf &= \landauh \\
  \veloTFC_{i;\ncoord;\ncoord}\atSurf + \veloTFC_{\ncoord;i;\ncoord}\atSurf &= \landauh \formPeriod \label{eqNavierBCAtS}
\end{align}
Note that the right identity of eq. \eqref{eqNormalVeloBCAtS} is achieved by using
$\veloTFC_{\ncoord}\big|_{\paraTF(\ncoord=-\frac{h}{2})} = \veloTFC_{\ncoord}\big|_{\paraTF(\ncoord=\frac{h}{2})} = \veloTFC_{\ncoord}\big|_{\paraTF(\ncoord=0)}
   = \normalVelocity $, the related second order difference quotient and $ \veloTFC_{\ncoord;\ncoord;\ncoord} = \partial_{\xi}^{2}\veloTFC_{\ncoord} $, since vanishing Christoffel symbols $ \GammaTF_{\ncoord\ncoord}^{K} $.

With all these tools from above, we are able to realize a thin film limit of eqs. \eqref{eqNSTF} -- \eqref{eqDirTF} for $ h \rightarrow 0 $, consistently,
by Taylor expansion of the equations at the surface.
The thin film $ \TFilm $ is a flat Riemannian manifold and therefore the Riemannian curvature tensor vanish, \ie\  covariant derivatives commute, 
and with the continuity equation \eqref{eqContiTF} we obtain 
\begin{align}
  \left[ \vecLaplace\veloTF \right]_{I} &= \left[ \divTF\nablaTF\veloTF + \nablaTF\divTF\veloTF \right]_{I}\notag\\ 
         &= \tensor{\veloTFC}{_{I}^{;K}_{;K}} + \tensor{\veloTFC}{^{K}_{;K;I}} 
          = \tensor{\veloTFC}{_{I}^{;K}_{;K}} + \tensor{\veloTFC}{^{K}_{;I;K}} \notag\\
         &= \left[ \divTF\Lie{\veloTF}\gTF \right]_{I}\formPeriod
\end{align}
Hence, we have to develop three divergence terms of 2-tensors at the surface in eqs. \eqref{eqNSTF} -- \eqref{eqDirTF},
namely $\divTF\arbTensorTF\atSurf$ for $\arbTensorTF$ being either $\Lie{\veloTF}\gTF$, $\ericksenStress$ or $\nablaTF\pTF$.
By using eqs. \eqref{eqNormalPBCAtS} -- \eqref{eqNavierBCAtS} it holds
\begin{equation}
	\label{eqArbTensorBCAtS}
	\begin{aligned}
		\arbTensorTFC_{i\ncoord}\atSurf = \landauh \formComma\quad &\partial_{\ncoord}\arbTensorTFC_{i\ncoord}\atSurf = \landauh \formComma \\
		\arbTensorTFC_{i\ncoord;\ncoord}\atSurf =& \landauh\formPeriod
	\end{aligned}
\end{equation}
The covariant tangential components of the divergence at the surface are
\begin{align*}
	\left[ \divTF\arbTensorTF \right]_{i}\AtSurf &= \tensor{\arbTensorTFC}{_{i}^{K}_{;K}}\AtSurf \\
	&= \left(\partial_{K}\tensor{T}{_{i}^{K}} - \GammaTF_{Ki}^{J}\tensor{T}{_{J}^{K}} +  \GammaTF_{KL}^{K}\tensor{T}{_{i}^{L}}\right)\AtSurf \formComma
\end{align*}
where it holds by using eq. \eqref{eqArbTensorBCAtS}
\begin{align}
	\partial_{K}\tensor{T}{_{i}^{K}}\AtSurf &= \partial_{k}\tensor{T}{_{i}^{k}}\AtSurf + \landauh \formComma\\
	\GammaTF_{Ki}^{J}\tensor{T}{_{J}^{K}}\AtSurf &= \left( \GammaS_{ki}^{j}\tensor{T}{_{j}^{k}} + \shapeOperatorc_{ik} \tensor{T}{_{\ncoord}^{k}}\right)\AtSurf + \landauh \formComma\\
	\GammaTF_{KL}^{K}\tensor{T}{_{i}^{L}}\AtSurf &= \GammaS_{kl}^{k}\tensor{T}{_{i}^{l}}\AtSurf + \landauh \formPeriod
\end{align}
Adding this up and take the metric compatibility of $ \GradSurf $ into account, we obtain
\begin{align}
	\label{eqDivArbTensor}
	\left[ \divTF\arbTensorTF \right]_{i}\AtSurf &= \left( \tensor{\arbTensorTFC}{_{i}^{k}}\AtSurf \right)_{|k} - \shapeOperatorc_{ik}\tensor{\arbTensorTFC}{_{\ncoord}^{k}}\AtSurf + \landauh \notag \\
	&= \gSC^{kl} \left( \left(\arbTensorTFC_{il}\AtSurf \right)_{|k} - \shapeOperatorc_{ik}\arbTensorTFC_{\ncoord l}\AtSurf \right) \notag \\
	&\qquad + \landauh \formPeriod
\end{align}
Note that all normal derivatives vanished here, \ie\ there is no need for a higher order expansion in $\ncoord$ of the thin film Christoffel symbols
as a consequence of the used boundary conditions.
To substantiate the tensor $ \arbTensorTF\in\Tangent^{(2)}\TFilm $, we first observe that
\begin{align}
	\veloTFC_{i;j}\AtSurf &= \left( \partial_{j}\veloTFC_{i} - \GammaTF_{ji}^{k}\veloTFC_{k} -  \GammaTF_{ji}^{\ncoord}\veloTFC_{\ncoord}\right)\AtSurf\notag\\ 
	&= \vbc_{i|j} - \normalVelocity\shapeOperatorc_{ij} \formComma \label{eqGradVeloTFAtS}\\
	\pTFC_{i;j}\AtSurf  &= \left( \partial_{j}\pTFC_{i} - \GammaTF_{ji}^{k}\pTFC_{k} -  \GammaTF_{ji}^{\ncoord}\pTFC_{\ncoord}\right)\AtSurf\notag\\
	&= \pbc_{i|j} + \landauh \formComma \label{eqGradPTFAtS} \\
	\pTFC_{\ncoord;j}\AtSurf &= \left( \partial_{j}\pTFC_{\ncoord} - \GammaTF_{j\ncoord}^{k}\pTFC_{k} \right)\AtSurf = \tensor{\shapeOperatorc}{_{j}^{k}}\pbc_{k} + \landauh \formComma
\end{align}
where $ \vbc_{i} := \veloTFC_{i}\atSurf $ and $ \pbc_{i} := \pTFC_{i}\atSurf $, 
\ie\ $ \vb := \ProjectSurf\veloTF\atSurf\in\Tangent^{1}\surf $ and in contravariant form $ \pb := \ProjectSurf\pTF\atSurf\in\Tangent^{1}\surf $.
We further obtain 
\begin{align*}
	\left[ \Lie{\veloTF}\gTF \right]_{il}\AtSurf &= \vbc_{i|l} + \vbc_{l|i} - 2\normalVelocity\shapeOperatorc_{il} \\
        &= \left[ \Lie{\vb}\gS - 2\normalVelocity\shapeOperatorc\right]_{il} \formComma\\
	\left[ \ericksenStress \right]_{il}\AtSurf &= \left[ \left(\nablaTF\pTF\right)^{T} \nablaTF\pTF \right]_{il}\AtSurf \\
	&= \left( \gTFC^{jk}\pbc_{j;i}\pbc_{k;l} + \pbc_{\ncoord;i}\pbc_{\ncoord;l} \right)\AtSurf \\
        &= \left[ \left(\GradSurf\pb\right)^{T} \GradSurf\pb + \left( \shapeOperator\pb \right) \otimes \left( \shapeOperator\pb \right) \right]_{il} + \landauh 
\end{align*}
which we define as the extrinsic surface Ericksen stress tensor $\surfaceEricksenStress:=\left(\GradSurf\pb\right)^{T} \GradSurf\pb + \left( \shapeOperator\pb \right) \otimes \left( \shapeOperator\pb \right)$, and finally get
\begin{align}
	\ProjectSurf\left(\vecLaplace\veloTF\right)\AtSurf &= \DivSurf\left( \Lie{\vb}\gS - 2\normalVelocity\shapeOperatorc \right) + \landauh \formComma \label{eqLaplaceVeloTFAtSI}\\
	\ProjectSurf\left(\vecLaplace\pTF\right)\AtSurf &= \laplaceDivGrad\pb - \shapeOperator^{2}\pb + \landauh \formComma \notag \\
	\ProjectSurf\divTF\ericksenStress\AtSurf &= \DivSurf\surfaceEricksenStress + \landauh \formComma\notag
\end{align}
where eq. \eqref{eqDivArbTensor} was used and $\laplaceDivGrad := \DivSurf\circ\GradSurf$ denotes the Bochner Laplacian.
Evaluating eq. \eqref{eqContiTF} at the surface and using the boundary condition \eqref{eqNormalVeloBCAtS} and the identity in eq. \eqref{eqGradVeloTFAtS} gives
\begin{align}
  0 &= \divTF\veloTF\AtSurf
     = \left( \gTFC^{ij} \veloTFC_{i;j} + \veloTFC_{\ncoord:\ncoord} \right)\AtSurf \notag\\
    &= \DivSurf\vb - \normalVelocity\meanCurvature + \landauh \formPeriod \label{eqContiTFAtS}
\end{align}
Furthermore, we introduce the two curl operators $\RotSurf: \Tangent^{(1)}\surf \rightarrow \Tangent^{0}\surf$ for vector fields and 
$\BigRotSurf: \Tangent^{0}\surf \rightarrow \Tangent^{(1)}\surf$ for scalar fields, see \cite{Nestleretal_JNS_2018} and \cite{Nitschkeetal_book_2017} for their definitions. 
Therefore, rewriting eq. \eqref{eqLaplaceVeloTFAtSI} yields
\begin{align}
	\label{eq:viscouse_alternative}
	\ProjectSurf\left(\vecLaplace\veloTF\right)\AtSurf &= \BigRotSurf\RotSurf\vb + 2\left(\gaussianCurvature\vb + \meanCurvature\GradSurf\normalVelocity \right) \notag \\
	&\qquad - 2\shapeOperator\GradSurf\normalVelocity + \landauh\notag \\
	&= -\laplaceDeRham\vb + 2\gaussianCurvature\vb + \GradSurf\left(\normalVelocity\meanCurvature\right) \notag \\
	&\qquad - 2\DivSurf\left(\normalVelocity\shapeOperator\right) +  \landauh\formComma
\end{align}
as a consequence of eq. \eqref{eqContiTFAtS} and the Weizenböck machinery, \ie\ interchanging covariant derivatives \wrt\ the Riemannian curvature tensor of $ \surf $, \cf \cite{Arroyoetal_PRE_2009}.
With
\begin{align*}
	\left\| \pTF \right\|^{2}_{\TFilm}\AtSurf &= \left( \gTFC^{ij}\pTFC_{i}\pTFC_{j} + \left( \pTFC_{\ncoord} \right)^{2} \right)\AtSurf = \left\| \pb \right\|^{2}_{\surf} + \landau\left( h^{4} \right)
\end{align*}
we get for the remaining terms on the right hand sides of eqs. \eqref{eqDirTF} and \eqref{eqNSTF}
\begin{align*}
	\ProjectSurf\left( \left( \left\| \pTF \right\|^{2}_{\TFilm} - 1 \right)\pTF \right)\AtSurf &= \left( \left\| \pb \right\|^{2}_{\surf} - 1 \right)\pb + \landau\left( h^{4} \right) \formComma \\
	\ProjectSurf\left( \nablaTF\pressTF \right)\AtSurf &= \GradSurf\pressure \formComma
\end{align*}
with the surface pressure $ \pressure := \pressTF\atSurf $.
Note that it holds for the relative velocity at the surface $\relveloTF\atSurf = \ProjectSurf\veloTF\atSurf$, \ie\ $U_{\ncoord}\atSurf = 0$ and $U^{i}\atSurf = \vbc^{i}$, on the left hand sides of eqs. \eqref{eqNSTF} and \eqref{eqDirTF}.
Hence, eqs. \eqref{eqGradVeloTFAtS} and \eqref{eqGradPTFAtS} now read 
\begin{align*}
  \left[ \dirnablaTF{\relveloTF}\veloTF \right]_{i}\AtSurf
      &= \vbc^{k}\veloTFC_{i;k}\AtSurf
       = \left[ \dirnablaS{\vb}\vb - \normalVelocity\shapeOperator\vb \right]_{i}\\
  \left[ \dirnablaTF{\relveloTF}\pTF \right]_{i}\AtSurf
      &= \vbc^{k}\pTFC_{i;k}\AtSurf
       = \left[ \dirnablaS{\vb}\pb \right]_{i} + \landauh \formPeriod
\end{align*}
By using $\partial_{t}\paraTF = \obveloTF = \obveloTFC_{\ncoord}\surfNormal$, $\obveloTFC_{\ncoord}\atSurf = \normalVelocity$ and
$\veloTF = \veloTFC^{k}\partial_{k}\paraTF + \veloTFC_{\ncoord}\surfNormal$ (analogously for $\pTF$) 
we obtain for the partial time derivatives 
\begin{align}
  \left[ \partial_{t}\veloTF \right]_{i}\AtSurf
      &= \left\langle \partial_{t}\veloTF, \partial_{i}\paraTF \right\rangle_{\TFilm}\AtSurf \notag\\
      &= \big\langle \left( \partial_{t}\veloTFC^{k} \right)\partial_{k}\paraTF 
                 + \veloTFC^{k}\partial_{k}\obveloTF \notag\\
      &\quad\quad+ \left( \partial_{t}\veloTFC_{\ncoord} \right)\surfNormal
                 + \veloTFC_{\ncoord} \partial_{t}\surfNormal               ,  \partial_{i}\paraTF \big\rangle_{\TFilm}\AtSurf\notag\\
      &= g_{ik}\partial_{t}\vbc^{k}
        + \vbc^{k}\left\langle \left(\partial_{k}\obveloTFC_{\ncoord} \right)\surfNormal + \normalVelocity\partial_{k}\surfNormal , \partial_{i}\paraTF \right\rangle_{\TFilm}\AtSurf\notag\\
      &\quad\quad - \normalVelocity \left\langle \surfNormal , \partial_{i}\obveloTF \right\rangle_{\TFilm}\AtSurf \notag\\
      &= g_{ik}\partial_{t}\vbc^{k} - \normalVelocity\left( \shapeOperatorc_{ik}\vbc^{k} + \partial_{i}\normalVelocity \right)
\end{align}
and
\begin{align}
   \left[ \partial_{t}\pTF \right]_{i}\AtSurf
      &= \left\langle \partial_{t}\pTF, \partial_{i}\paraTF \right\rangle_{\TFilm}\AtSurf \notag\\
      &= \left\langle \left( \partial_{t}\pTFC^{k} \right)\partial_{k}\paraTF + \pTFC^{k}\partial_{k}\obveloTF,  \partial_{i}\paraTF \right\rangle_{\TFilm}\AtSurf \notag \\
      &\qquad + \landauh \notag\\
      &= g_{ik}\partial_{t}\pbc^{k} - \normalVelocity\shapeOperatorc_{ik}\pbc^{k} + \landauh \formPeriod
\end{align}
Therefore, we have
\begin{align}
	\label{eqAccelTFAtS}
	\ProjectSurf\left[ \partial_{t}\veloTF + \dirnablaTF{\relveloTF}\veloTF \right]\AtSurf &= \left( \partial_{t}\vb^{i} \right)\partial_{i}\paraS + \dirnablaS{\vb}\vb \notag \\ 
	&\qquad- \normalVelocity\left( 2\shapeOperator\vb + \GradSurf\normalVelocity \right)
\end{align}
for the tangential part of the fluid acceleration in eq. \eqref{eqNSTF}.
The same term was proposed in \cite{Yavarietal_JNS_2016} by variation of the kinetic energy of a moving manifold in the context of Lagrangian field theory.
Moreover, we also find this acceleration term in \cite{NitschkeVoigt_2019}, where a covariant material derivative is derived in terms of covariant tensor transport 
through a three dimensional moving spacetime embedded in a four dimensional absolute space. 
In this context eq. \eqref{eqAccelTFAtS} can be obtained by taking the spatial part of the covariant material derivative for the special case of velocity fields and a transversal observer.
Evaluating the transport term for the director field $ \pTF $ at the surface yields
\begin{align*}
	\ProjectSurf\left[ \partial_{t}\pTF + \dirnablaTF{\relveloTF}\pTF \right]\AtSurf = \left( \partial_{t}\pb^{i} \right)\partial_{i}\paraS + \dirnablaS{\vb}\pb - \normalVelocity\shapeOperatorc\pb\notag \formComma
\end{align*}
which can also be found in \cite{NitschkeVoigt_2019}, but as the spatial part of the covariant material derivative of a so-called instantaneous vector field from a point of view of a transversal observer.
Finally, under boundary conditions \eqref{eqNormalPBC} -- \eqref{eqNavierBC} and $ h\rightarrow 0$, eq. \eqref{eqContiTF} 
and the tangential parts of eqs. \eqref{eqNSTF} and  \eqref{eqDirTF} reduces to 
\begin{align}
	&\left( \partial_{t}\vb^{i} \right)\partial_{i}\paraS + \dirnablaS{\vb}\vb - \normalVelocity\left( 2\shapeOperator\vb + \GradSurf\normalVelocity \right) \notag \\
	&\phantom{\DivSurf\vb}= \viscosity\left( -\laplaceDeRham\vb + 2\gaussianCurvature\vb + \GradSurf\left(\normalVelocity\meanCurvature\right) - 2\DivSurf\left(\normalVelocity\shapeOperator\right) \right)\notag \\
	&\phantom{\DivSurf\vb}\qquad -\GradSurf\pressure - \lambda\DivSurf\surfaceEricksenStress \label{eqNSS} \\
	\label{eqContiS}
	&\DivSurf\vb = \normalVelocity\meanCurvature\\
	\label{eqDirS}
	&\left( \partial_{t}\pb^{i} \right)\partial_{i}\paraS + \dirnablaS{\vb}\pb - \normalVelocity\shapeOperatorc\pb \notag\\
	&\phantom{\DivSurf\vb}= \K\left( \laplaceDivGrad\pb - \shapeOperator^{2}\pb \right) - \Kn\left( \left\| \pb \right\|_{\surf}^{2} - 1 \right)\pb
\end{align}
in $ \surf\times\R_{+} $.
This system of PDEs has full rank, \ie\ it contains five independent coupled equations with five degree of freedoms 
$\vbc^{1}$, $\vbc^{2}$, $\pbc^{1}$, $\pbc^{2}$, depending on an arbitrary choice of local coordinates, and $\pressure$.
A full discussion about the normal parts of eqs. \eqref{eqNSTF} and \eqref{eqDirTF} does not belong to this paper. 
Nevertheless, the normal parts would give us two additional equations, consistently \wrt\ $h$, and
two new free scalar valued quantities $ \gamma_1,\gamma_2\in\Tangent^{0}\surf $.
With our boundary conditions and assumption, \wrt\ the moving thin film geometry, 
this would be $ \gamma_1 := \partial_{\ncoord}\pressTF\atSurf $ and $ \gamma_2:= \pTFC_{\ncoord;\ncoord;\ncoord}\atSurf = \partial_{\ncoord}^{2}\pTFC_{\ncoord}\atSurf $.
Both degrees of freedom would occur as zero order differential terms, \ie\ the upcoming normal equations would not have any influence to eqs. \eqref{eqNSS} -- \eqref{eqDirS}.
Therefore, the thin film limit of the normal part equations can be omitted as long as we are not interested in the quantities $\gamma_1$ and $\gamma_2$.

The partial time derivatives in eqs. \eqref{eqNSS} and \eqref{eqDirS} are realized only at the contravariant vector proxies of $\vb$ and  $\pb$ 
\wrt\ locally defined charts at the surface.
The reason for the absence of an intrinsic covariant vector operator notation for the time derivative (similar to $\GradSurf$) is that the time $t$ is not a coordinate of a
moving space in a pure spatial perspective, especially for a moving surface with $\normalVelocity\neq 0$.
Unfortunately, most of the numerical tools for solving surface PDEs do not work with a locally defined vector basis. 
They mimic vector-valued problems as a system of scalar-valued problems under the assumption of Euclidean coordinates.
This means for the surface problem \eqref{eqNSS} -- \eqref{eqDirS} that the tangential velocity field and the director field are considered to be vector fields in $\R^{3}$, which results in an under-determined problem. The two additional degrees of freedom can be handled in different ways, \eg\ by penalty methods \cite{Nestleretal_JNS_2018,Reutheretal_PF_2018} or by using Lagrange multipliers \cite{Jankuhnetal_preprint_2017}.
The terms $\partial_{t}\vb$ and $\partial_{t}\pb$ makes certainly sense, if we consider $\vb,\pb\in\Tangent\R^{3}$,
but note that in general $\partial_{t}\vb$ as well as $\partial_{t}\pb$ are no longer part of the tangential space of the surface $\surf$.
Nevertheless, we can use only the tangential part of 
$\partial_{t}\vb = \left( \partial_{t}\vbc^{j} \right)\partial_{j}\paraS + \vbc^{j}\partial_{j}\partial_{t}\paraS$ for a transversal observer, \ie\
\begin{align*}
	\left[ \partial_{t}\vb  \right]_{i} &= \left\langle \partial_{t}\vb, \partial_{i}\paraS \right\rangle_{\R^{3}} = \gSC_{ij}\partial_{t}\vbc^{j} - \normalVelocity\shapeOperatorc_{ij}\vbc^{j} \formPeriod
\end{align*}
Analogously, the same holds for $\partial_{t}\pb$.
Finally, we obtain by rewriting eqs. \eqref{eqNSS} -- \eqref{eqDirS}
\begin{align}
	\label{eqNSSAlt}
	&\ProjectSurf\partial_{t}\vb + \dirnablaS{\vb}\vb - \normalVelocity\left( \shapeOperator\vb + \GradSurf\normalVelocity \right) \notag\\
	&\phantom{\DivSurf\vb}= \viscosity\left( -\laplaceDeRham\vb + 2\gaussianCurvature\vb + \GradSurf\left(\normalVelocity\meanCurvature\right) - 2\DivSurf\left(\normalVelocity\shapeOperator\right) \right)\notag \\
	&\phantom{\DivSurf\vb}\quad -\GradSurf\pressure - \lambda\DivSurf\surfaceEricksenStress \\
	\label{eqContiSAlt}
	&\DivSurf\vb = \normalVelocity\meanCurvature\\
	\label{eqDirSAlt}
	&\ProjectSurf\partial_{t}\pb + \dirnablaS{\vb}\pb \notag \\
	&\phantom{\DivSurf\vb} = \K\left( \laplaceDivGrad\pb - \shapeOperator^{2}\pb \right) -\Kn\left( \left\| \pb \right\|_{\surf}^{2} - 1 \right)\pb
\end{align}
in $ \surf\times\R_{+} $, if $ \vb,\pb\in\Tangent\surf\subset\Tangent\R^{3} $.

Eq. \eqref{eqDirTF} together with the boundary condition \eqref{eqNeumannBC} is the $L_{2}$-gradient flow along the material motion to minimize the Frank-Oseen energy functional with material constants $\K = K_{11} = K_{22} = K_{33}$ and $K_{24} = 0$, see \cite{Morietal_JJAP_1999}. 
In the thin film limit, this leads to the minimization of the surface Frank-Oseen energy $\frac{\K}{2}\int_{\surf}\left\| \GradSurf\pb \right\|^{2}_{\surf}\dS$.
This situation differs from \cite{Nestleretal_JNS_2018}, where the one-constant approximation $\K = K_{11} = K_{22} = K_{33} = -K_{24}$ was assumed, which leads to minimizing the distortion energy $\frac{\K}{2}\int_{\surf}(\RotSurf\pb)^{2} + (\DivSurf\pb)^{2}\dS$. 
However, for the case $\Kn\rightarrow\infty$, where $\left\| \pb \right\|^{2}_{\surf} = 1$ a.\,e., both energies only differ 
by a constant value $\frac{K}{2}\int_{\surf} \gaussianCurvature \dS = \pi K\chi\left( \surf \right)$, where $\chi\left( \surf \right)$ denotes the Euler characteristic.
Thus, the minimizers of both energies are equal.

\bibliography{references.bib}

\end{document}